\documentclass[lettersize,journal]{IEEEtran}

\usepackage{amsmath,amsfonts}
\usepackage{algorithmic}
\usepackage{algorithm}
\usepackage{array}
\usepackage[caption=false,font=normalsize,labelfont=sf,textfont=sf]{subfig}
\usepackage{textcomp}
\usepackage{stfloats}
\usepackage{url}
\usepackage{verbatim}
\usepackage{graphicx}
\usepackage{cite}
\usepackage{multirow}
\usepackage{epstopdf}
\usepackage[utf8]{inputenc}
\usepackage{float}
\usepackage{caption} 
\usepackage{authblk}
\usepackage{bm}
\usepackage[font=small]{caption}
\usepackage{xcolor}
\usepackage{hyperref}
\usepackage{amsmath}
\usepackage{makecell}

\abovedisplayshortskip
\belowdisplayshortskip

\DeclareGraphicsExtensions{.jpg,.png,.pdf}
\begin{document}

\title{Multi-BD Symbiotic Radio-Aided 6G IoT Network: Energy Consumption Optimization with QoS Constraint Approach}
\author{Rahman Saadat Yeganeh\,$^1$, Mohammad Javad Omidi\,$^{1,2}$ and Mohammad Ghavami\,$^3$, ~\IEEEmembership{Senior Member, IEEE}
\thanks{$^1$Department of Electrical and Computer Engineering, Isfahan University of Technology, Isfahan 84156-83111, IRAN (emails: r.saadat@ec.iut.ac.ir, omidi@iut.ac.ir).}
\thanks{$^2$Department of Electronics and Communication Engineering, Kuwait College of Science and Technology (KCST), Doha 35003, Kuwait}
\thanks{$^3$Electrical and Electronic Engineering Department, London South Bank University, London SE1 0AA, U.K.(email: ghavamim@lsbu.ac.uk).}}
\maketitle

\begin{abstract}
The commensal symbiotic radio (CSR) system is proposed as a novel solution for connecting systems through green communication networks. This system enables us to establish secure, ubiquitous, and unlimited connectivity, which is a goal of 6G. The base station uses MIMO antennas to transmit its signal. Passive IoT devices, called symbiotic backscatter devices (SBDs), receive the signal and use it to charge their power supply. When the SBDs have data to transmit, they modulate the information onto the received ambient RF signal and send it to the symbiotic user equipment, which is a typical active device. The main purpose is to enhance energy efficiency in this network by minimizing energy consumption (EC) while ensuring the minimum required throughput for SBDs. To achieve this, we propose a new scheduling scheme called Timing-SR that optimally allocates resources to SBDs. The main optimization problem involves non-convex objective functions and constraints. To solve this, we use mathematical techniques and introduce a new approach called sequential quadratic and conic quadratic representation to relax and discipline the problem, leading to reducing its complexity and convergence time. The simulation results demonstrate that the proposed approach outperforms other outlined schemes in reducing EC.
\end{abstract}
\section{Introduction}
\subsection{Background}
Future networks such as B5G and 6G must be capable of covering billions of internet-of-things (IoT) and wireless devices\cite{8879484}. The most critical challenges in this regard include ensuring adequate energy supply, providing the desired frequency spectrum, and enabling devices to operate independently of communication infrastructure\cite{r1,r3,r28}. The energy efficiency (EE) and spectral efficiency (SE) can be improved and a dense network can be managed without the need for additional EC or heavy processing in the central core of the network by addressing and overcoming these challenges\cite{r29}. 

To address these challenges, researchers have proposed various solutions; Cognitive radio (CR) can facilitate dynamic spectrum allocation, allowing the secondary user to access the spectrum allocated to the primary user in an opportunistic or spectrum sharing manner, thereby meeting their communication requirements \cite{r22,r23,r24}. Wireless energy transfer (WET) techniques can provide sustainable and efficient power to wireless devices\cite{r11}. Another promising approach is ambient backscatter communication, which allows passive backscatter devices (BDs) to modulate their information onto ambient RF signals (such as cellular, TV, or WiFi signals) and communicate with each other. BDs can also harvest energy from the incident signals in the environment, making it an attractive solution for low-power and low-cost communication \cite{r19,r26}. 

However, each of these solutions comes with its own limitations alongside their advantages. For example, in CR, there is a need for complete information about the power, spectrum and duration of its use by the primary transmitter, accurate channel state information (CSI), and synchronization of the primary and secondary transmitters with each other.  As a result, these requirements can only be fulfilled by active devices, which increases EC, reduces battery life, and limits the development of future wireless network services \cite{r25}. WET is exclusively used for energy supply applications, but it faces its own set of challenges, such as the doubly near far problem \cite{Qiang,r14}. AmBC also faces several challenges, including non-cooperation in information transmission between BS and BDs, leading to interference that makes it difficult for the receiver to jointly decode their signals. Furthermore, the performance of AmBC is not always stable due to the possibility of changes in the BS signal or its location \cite{r27}.

Based on the above explanations, we need a comprehensive solution that not only takes advantage of each of the above methods but also eliminates their disadvantages. The new technology that possesses this feature is Symbiotic Radio, which is currently one of the fascinating topics in the scientific and industrial fields \cite{r20,r21}. 

The SR network can be classified into parasitic SR (PSR) and commensal SR (CSR) based on the relationship between the symbol periods of the symbiotic BDs (SBDs) and the BS \cite{8907447}. In PSR setup, SBDs can exchange information at a high rate, but it also suffers from interference between the signals of SBDs and BS in the receiver, necessitating complex interference cancellation techniques. Furthermore, PSR requires synchronization between the BS, BD, and receiver. On the other hand, CSR is suitable for IoT networks with low data rates, and addresses the drawbacks of the PSR system. By reducing interference between different network components, the receiver can perform joint decoding of information from both the BS and SBDs, enabled by transmit collaboration between them \cite{r22,r25}.

To ensure that the SR networks can accommodate a large number of devices, it's important to design the network in a way that supports multiple SBDs. In doing so, challenges arise, such as the possibility of minor user interference if multiple access schemes are not appropriately designed. On the other hand, given the large number of users in the network, optimizing EC is vital for maintaining network stability and achieving the desired minimum quality of service (QoS) for SBDs. Furthermore, reducing energy consumption has significant environmental benefits, making it an essential aspect of green communication and facilitating the easier implementation of self-sustaining networks \cite{r8,r10,8922617}. To achieve these goals, some solutions have been proposed.

The articles \cite{8907447,13691370}, proposes a novel SR technique for passive IoT devices. This approach involves integrating a BD with a primary communication system, and designing a  primary transmitter and receiver to optimize both the primary and BD transmissions. The decoding strategy used in the receiver is based on successive interference cancellation (SIC).
In \cite{r38}, \cite{9201474}, the authors aimed to achieve the maximum EE by efficiently allocating resources while ensuring QoS requirements in NOMA-backscatter communication networks and NOMA-heterogeneous networks, respectively.
Also, the authors in \cite{r39} utilize the dinkelbach algorithm to solve the problem of maximizing EE in the SR system with multiple BDs that can harvest energy from ambient signals. The scheduling protocol employed in this system is time division multiple access (TDMA), which allows BDs to modulate their information on the ambient signal without interfering with each other, as they take turns to transmit. 

Due to the fact that the TDMA technique does not support concurrent transmissions by BDs, leading to a reduction in both EE and SE in the network, the authors in \cite{rn98,r40,turbocharging,r41,RN111} proposed a technique for enabling concurrent transmission by BDs in the symbiotic communication model.
The article \cite{rn98,yeganeh2023energy} discusses the random distribution of SBDs in the network, which causes signals to reach the receiver at different power levels. To tackle this, a SIC process is employed to eliminate interference from signals with higher power levels than the desired SBD signal. Also, \cite{r40}, the authors present a technique for preventing interference among multiple reflected signals in the symbiotic communication model. They use a simple coding algorithm that does not require strict synchronization and passive users use mutually orthogonal chips for encoding, eliminating interference through orthogonal interference and enabling concurrent transmission for multiple users. In addition, the paper \cite{turbocharging} proposes a low-power encoding technique called $\mu$code to enhance the communication range and enable concurrent transmissions in AmBC. $\mu$code employs a form of code division multiple access (CDMA), where each message is assigned a single randomly generated code that is orthogonal to all previously generated codes. This technique effectively reduces interference and allows for concurrent transmissions. Furthermore, article \cite{r41} addresses a multi-BD SR system, where a cooperative receiver can simultaneously receive and detect the data from the primary transmitter and multiple BDs. The authors assume a random code-assisted multiple access scheme for the multiple BDs, where the transmit power of the primary transmitter and the reflection coefficients of the BDs are jointly optimized. In this scheme, each BD selects its non-orthogonal random code to backscatter its information instantly. In \cite{RN111}, the authors introduce a sparse-coded AmBC scheme for achieving massive IoT connectivity. This scheme allows multiple devices to transmit their information concurrently, which reduces interference and improves the overall network performance.

In this paper, we focus on enhancing the EE of SR networks with the CSR setup in scenarios involving multiple SBDs. SBDs are passive IoT sensors that rely on harvesting energy from ambient signals and have limited power supplies \cite{r36}. These devices collect and transmit information to the destination once they have enough energy and detect changes in their environment. In our system model, we aim to minimize the network EC while guaranteeing a minimum required throughput for multiple randomly distributed IoT devices (SBDs). The major contributions of this paper are summarized as follows:
\begin{itemize}
\item{First, we present a novel SR system that utilizes multiple SBDs in a CSR setup, which allows for the full exploitation of the mutualism relationship of SR. Specifically, our model enables SBDs to transmit their information to their intended destination as soon as they have data to send. To ensure a high-quality user experience, we also incorporate a QoS constraint in the system, which guarantees a minimum transmission rate for SBDs.}
\item{Second, to optimize the allocation of resources between SBDs, we propose a multiple access technique called timing-SR (T-SR). The T-SR model utilizes a two-mode variable time slot consisting of energy harvesting and environment sensing (EHS), and modulation and transmission of information (MTI) modes. This approach allows users to harvest energy as needed for their data transmissions and reduces EC by eliminating idle modes. Also, they can send their information in continuous time slots without interfering with other SBDs.}
\item{Third, the objective of the proposed model is to minimize EC in a network while ensuring that the SBDs meet their required throughput and energy harvesting targets. Since this is a non-convex optimization problem, we employ mathematical techniques such as semidefinite programming (SDP) and the difference of two convex function methods to solve it. Additionally, we introduce a novel approach that called sequential quadratic (SQ) and conic quadratic representation (CQR) to relax and discipline the optimization problem and reduce its computational complexity. By utilizing these techniques, we can efficiently solve the optimization problem, which makes it suitable for fast coverage of large-scale networks.}
\item{Finally, in the simulation section, we compare the EC of the proposed SR system using the T-SR scheduling mode with the conventional TDMA scheme in the SR network. Moreover, we evaluate the EE of the proposed model by comparing it with other well-established IoT protocols, thereby showcasing its effectiveness in dense networks like B5G and 6G. Our findings reveal that the SR system with the T-SR scheduling mode is a highly promising solution for maximizing energy efficiency in such networks.}
\end{itemize}

This paper is structured as follows. In Section II, we present the proposed system model for the SR. Section III focuses on the EC problem and considers the minimum user throughput requirement. We introduce the SQ and CQR methods in this section and use them to solve the main problem. In Section IV, we analyze the computational complexity of the proposed methods. In Section V, we validate our analytical findings through simulations and comparisons with other work. Finally, in Section VI, we summarize our conclusions and discuss future work.

\textbf{\emph{Notations}}: $\left\langle {a,b} \right\rangle $ denotes the inner product of
$a$ and $b$, $Tr\left( {\bf{A}} \right)$, ${{\bf{A}}^H}$, ${{\bf{A}}^T}$, $\left\| {\bf{A}} \right\|$ 
  denote the trace, conjugate transpose, transpose, and norm of the matrix ${\bf{A}}$, respectively. The positive semi-definite was denoted as  ${\mathbf{A}}\succeq 0$  and $\nabla $  shows the gradient operator.

\section{SYSTEM MODEL}
As illustrated in the system model of Fig.1, in the symbiotic radio (SR) system, the base stations (BS) equipped with 
${N}$ array antennas, a single symbiotic user equipment (SUE) is capable of functioning as a highly powerful active device, utilizing a single antenna, and $I$ single antenna symbiotic backscatter devices (SBDs) are considered. The SBDs are randomly distributed in the network with the different distance to SUE and harvest energy from the ambient signal transmitted by BS. The proposed SR system operates in two main phases. During the first phase, SBDs harvest energy from ambient signals. In this phase, SBDs, as IoT devices, can constantly sense the environment using embedded sensors (such as temperature, humidity, etc.); hence, this phase is called energy harvesting and environment sensing (EHS). In the second phase, the SBDs modulate their information on the received BS signal and backscatter the information signal to the SUE. This phase is called the modulation and transmission of information (MTI) by SBDs.
\begin{figure}
\centering
\includegraphics[width=8.7cm]{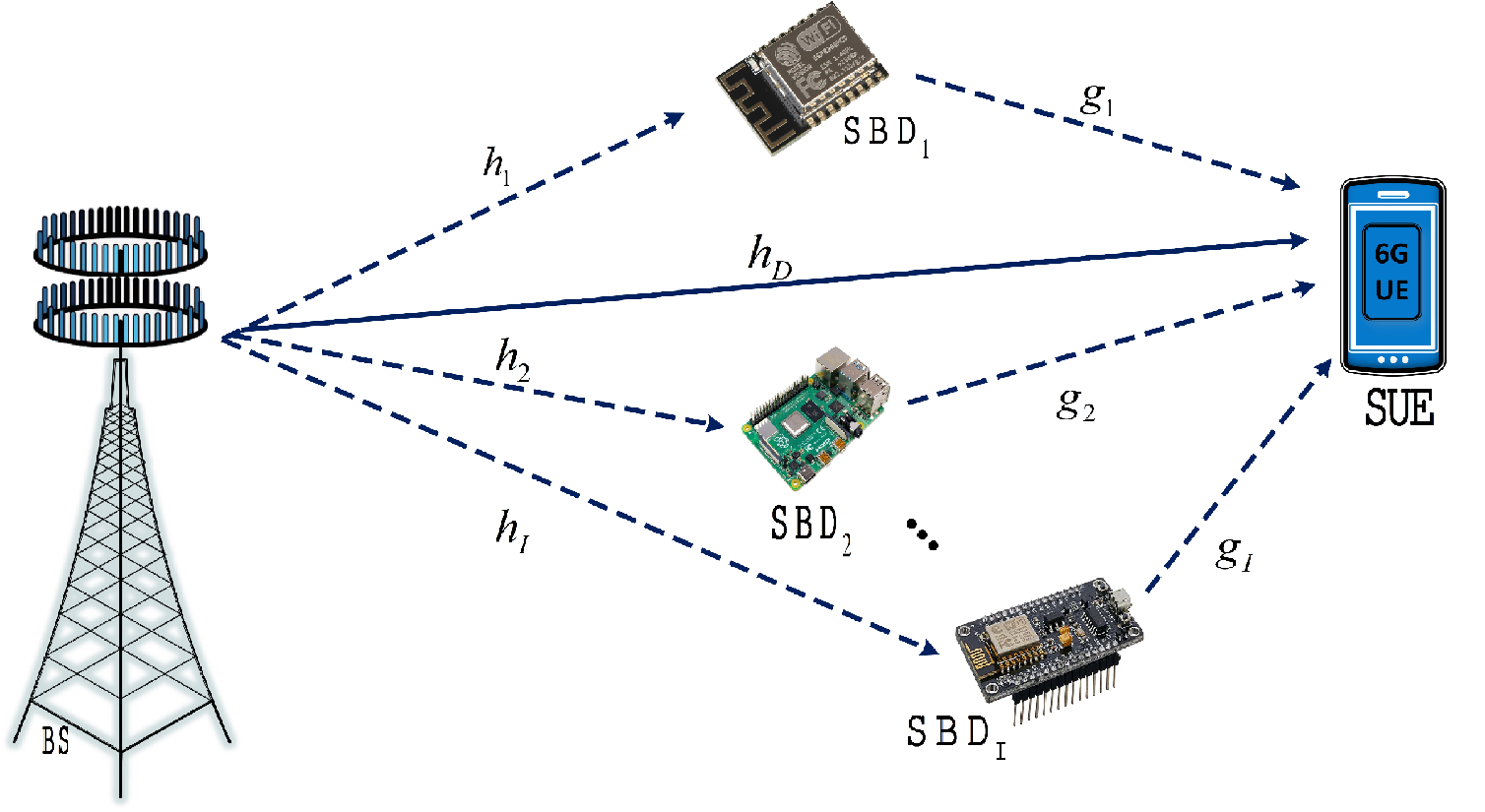}
\caption{Symbiotic radio system model with multiple SBDs } 
\end{figure} 

\begin{figure}
\begin{center}
\includegraphics[width=8.9cm]{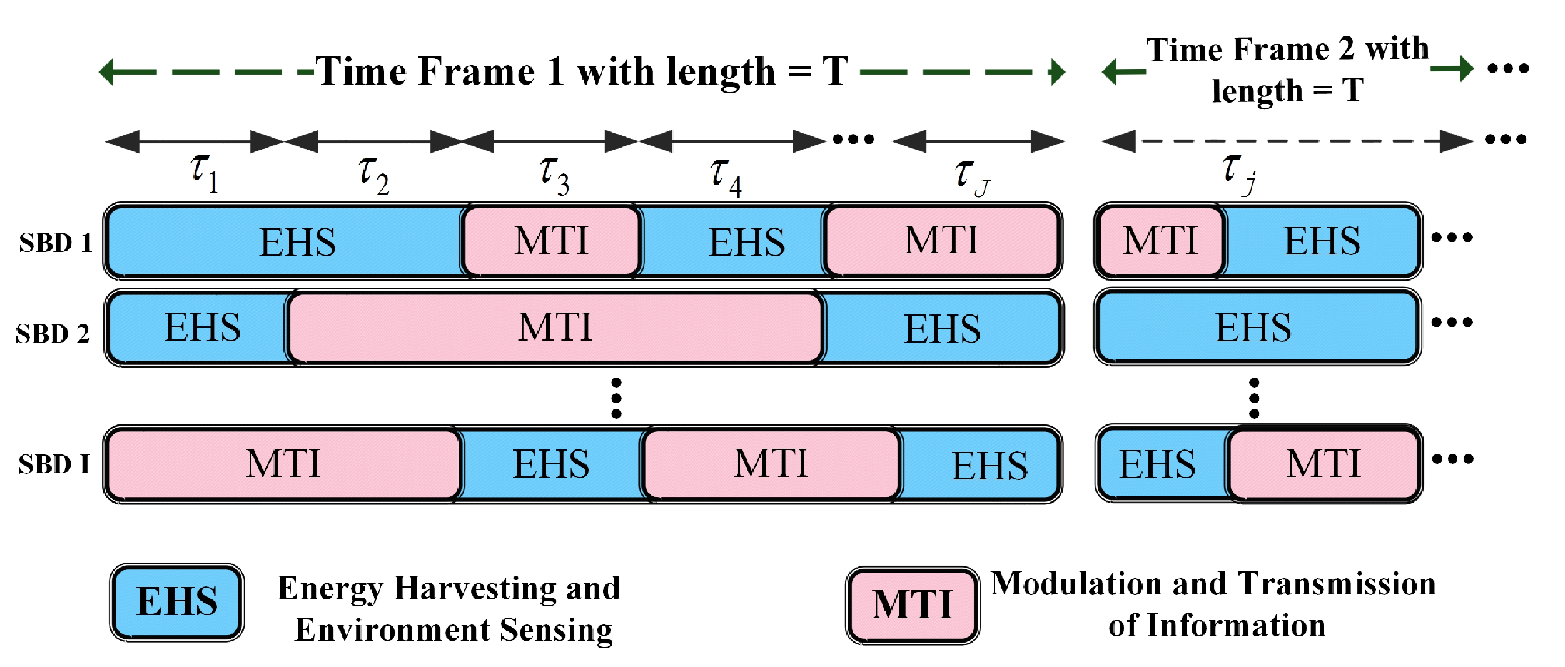}
\caption{The TDD frame for EHS and MTI modes in T-SR instantaneous transmission model } 
\end{center}
\end{figure}

The SBDs receive BS signal within ${\tau _j},j = 1,2,...,J$ time slots, where $J$ is the total number of time slots allocated to SBDs on the network, assumed to be equal to $I$. In the proposed SR system, time slots are identical, and SBDs can use one or more slots in each frame for MTI or EHS functions, with the number of slots required for transmission being proportional to the length of information and the energy required for it. Fig. 2 illustrates the scheduling of energy harvesting and data transmission between SBDs and SUEs in the proposed SR system, which uses a time division duplexing (TDD) scheme known as T-SR mode. Under this mode, SBDs initiate data transmission at the beginning of a time slot within each time frame of duration $T$, and can continue until the end of that slot or the next one.

We define $\upsilon_i$ as the set of time slots used for data transmission related to ${\text{SB}}{{\text{D}}_{\text{i}}}$. As an example, in Fig. 2, 
${\text{SB}}{{\text{D}}_{\text{2}}}$ transmits data during the 
${\tau _2}$,
 ${\tau _3}$, and ${\tau _4}$ time slots, so we have ${{\upsilon_2}}  = \left\{ {2,3,4} \right\}$. Therefore, in the T-SR model, we can express the time slots as:
\begin{equation}
{\tau _{j = 1,2,...,J}} = \left\{ {\begin{array}{*{20}{c}}
{{\rm{MTI}}}&{j \in {{\upsilon_i }}}\\
{{\rm{EHS}}}&{j \notin \upsilon_i }
\end{array}} \right.
\end{equation}

It is assumed that all SBDs have sufficient initial charges. Due to the single antenna constraint, SBDs can either harvest energy or transmit signals at a given time. Therefore, during transmission, SBDs switch from MTI mode to EHS mode whenever their batteries need to be recharged. The T-SR mode in the SR network is utilized to enhance both EE and throughput. Furthermore, there is no need for complex processing on the network core to reallocate frequencies to SBDs, as they use ambient signal frequencies. In this paper, we aim to investigate the EC of the SR system and compare the EC of the T-SR scheduling method with other technique in the simulation section.
\subsection{Problem Formulation}
 \emph{\textbf{Signal Model at ${\text{SB}}{{\text{D}}_{\text{i}}}$}}: In Fig.1, we assume that the BS with ${N}$ antennas and SBDs and SUE are single antenna. The signal transmitted by  $n$-th antenna on BS to SBDs in  $j$-th time slot is denoted by ${x_{n,j}}$ with zero mean and ${\mathbb{E}}\left[ {{{\left| {{x_{n,j}}} \right|}^2}} \right]={p_{BS}}$  which is the transmit power of  $n$-th antenna. In this paper, ${p_{BS}}$ is considered in 
${x_{n,j}}$ and it is not shown in relations. The ambient signal vector at the time slot ${\tau _j}$
 is defined as ${{\bf{x}}_j} \buildrel \Delta \over = {\left[ {{x_{1,j}},{x_{2,j}},...,{x_{N,j}}} \right]^T}$  and the channel vector from  $n$-th antenna to ${\text{SB}}{{\text{D}}_{\text{i}}}$ 
 is modeled by $h_{n,i}^{}$
 . Hence, the complex channel vector for ${\text{SB}}{{\text{D}}_{\text{i}}}$
 can be represented as ${{\bf{h}}_i} \buildrel \Delta \over = {\left[ {{h_{1,i}},{h_{2,i}},...,{h_{N,i}}} \right]^T}$, where the channel experiences flat fading and remains constant within a single time frame. It's worth noting that the channel state information (CSI) is readily available in all scenarios.
  Therefore, the received signal in the  ${\text{SB}}{{\text{D}}_{\text{i}}}$  at the time slot ${\tau _j},{\forall j}$  is: 
\begin{equation}
{y_{i,j}} = {\bf{h}}_i^H{{\bf{x}}_j} + {n_i}\,\,,\,\,\,\,i,j \in \bf{\Psi} 
\end{equation}
where ${{\bf{\Psi }} \buildrel \Delta \over = \left[ {1,2,...,I} \right]}$ and  ${{n_i} \sim \mathcal{CN}(0,{\sigma _i}^2)}$ 
is the circularly symmetric complex Gaussian (CSCG) distribution and it is assumed to be independent of the signal ${{{\bf{x}}_j}}$.
Let us define ${\varepsilon _{ij}}$ as the energy harvested by ${\text{SB}}{{\text{D}}_{\text{i}}}$ in ${\tau_ j},{j \notin \upsilon_i }$. According to Eq. (2), the maximum energy that can be harvested by ${\text{SB}}{{\text{D}}_{\text{i}}}$ in ${\tau _j},{j \notin \upsilon_i }$, occurs, when the power reflection coefficient ${p_i}$ (for ${\tau _{j \in \upsilon_i }}$ ,\({\left| {{p_i}} \right|^2} \le 1\)) is equal to zero. Thus, we can express the relation as follows:
 {\small\begin{equation}
\begin{array}{l}
{\varepsilon _{ij}} \leq {\eta _i}{(1-{p_i})}{\tau _{j \notin \upsilon_i}}\mathbb{E}\left[ {{{\left| {{y_{i,j}}} \right|}^2}} \right]  \approx {\eta _i}{\tau _{j \notin \upsilon_i}}{\bf{x}}_j^H{{\bf{h}}_i}{\bf{h}}_i^H{{\bf{x}}_j}\\
 \,\,\,\,\,\,\,\,\,\,\,\,\,\,\,\,\,\,\,\,\,\,\,\,\,\,\,\,\,\,\,\,\,\,\,\,\,\,\,\,\,\,\,\,\,\,\,\,\,\,\,\,\,\,\,\,\,\,\,\,\,\,\,\,\,\,\,\,\,\,\,\,\,\,\,\,\,\,\,\,\,\,\,\,\,\,\,\,\,\,\,\,\,\,\,\,\,\,\,\, , i \in {\bf{\Psi }},{j \notin \upsilon_i }
\end{array}
\end{equation}}
where, \(0 \le {\eta _i} \le 1\)  is the  energy conversion efficiency (for ${\tau _{j \notin \upsilon_i }}$) by  ${\text{SB}}{{\text{D}}_{\text{i}}}$.

The total energy sent by BS is the sum of energy sent by each of antennas (${{\rm{E}}_{\rm{T}}}$), so:
 	{\small\begin{equation}{{\rm{E}}_{\rm{T}}} = \sum\limits_{n = 1}^{N} {E_{{T_{BS}}}^n} \end{equation}}

${{\rm{E}}_{\rm{T}}}$ can also be expressed as the sum of the energy of transmitted signals by BS at all times during a frame.
    {\small  \begin{equation}{\rm{E}}_{\rm{T}}^{} = \sum\limits_{j = 1}^J {{\tau _j}{\bf{x}}_j^H{\bf{x}}_j^{}} \,\,,\,\,\,\,\,{\forall j} \end{equation}}

The main purpose of this paper is to minimize  ${{\rm{E}}_{\rm{T}}}$ to increase EE in the SR system while satisfying the minimum required data rate of the SBDs.

 \emph{\textbf{Signal Model at ${\text{SUE}}$}}: We consider the commensal symbiotic radio (CSR) setup, in which, for each transmission of SBD data symbol, $K$(  $k=1,2,...,K$, $K>>1$) symbols are transmitted by the BS. The information signal of ${\text{SBD}_{\text{i}}}$ is denoted by ${{\bf{s}}_{\bf{i}}} \buildrel \Delta \over = \left[ {{s_{i,a}},{s_{i,b}},...,{s_{i,z}}} \right]_{\left\{ {a,b,...,z} \right\} \in \upsilon_i }^T$. This signal should be modulated on the RF wave of the BS signal (${y_{i,j}}$), and then transmitted to the SUE during the MTI time slots ($\upsilon_i $). This can be done using various modulation techniques such as ASK, PSK, and so on.
Therefore, the received signal at SUE, transmitted by $k$-th symbol of the BS (through the direct link) and ${\text{SB}}{{\text{D}}_{\text{i}}}$ (through the backscatter link), in the time slot ${\tau _j},{j \in \upsilon_i}$ is as follows:
 {\small\begin{equation}
\begin{array}{l}
 y_{i,(j \in \upsilon_i) }^{UE}(k) = \sqrt {{p _i}}{}{\bf{h}}_i^H{{\bf{x}}_j}(k){g_i}{s_{i,(j \in \upsilon_i) }} + \\
 \,\,\,\,\,\,\,\,\,\,\,\,\,\,\,\,\,\,\,\,\,\,\,\,\,\,\,\,\,\,\,\,\,\,\,\, \sqrt {{p _i}}{g_i}{s_{i,(j \in \upsilon_i )}}{n_i} + {\bf{h}}_D^H{{\bf{x}}_j}(k) + {I_{0}} + {n_{UE}}
\end{array}
\end{equation}}
where ${g_i}$ is the complex channel gain from ${\text{SB}}{{\text{D}}_{\text{i}}}$ to SUE, and \({n_{UE}} \sim \mathcal{CN}(0,{\sigma _{UE}}^2)\)  is the CSCG distribution at the SUE. Also, ${I_{0}}$ is the possibility of interference caused by signals sent by other SBDs.
Note that since SBDs have a passive components, they receive little noise. Also, the power of the second term of Eq.(6) is much smaller than of the ${n_{UE}}$ due to the pathloss link. So, the noise of SBDs can be negligible when backscattering data \cite{r41,r48}.

 \emph{\textbf{Interference Model in The SR Network}}: The third term in $y_{i,(j \in \upsilon_i)}^{UE}$ of Eq.(6), refers to the ambient signal sent from the BS directly to the SUE. This signal can interfere with the desired signal transmitted by the SBDs. Due to the double fading effect, the power level of the SBDs signal is lower than that of the ambient signal. To remove the ambient signal, advanced techniques such as the ML-detector \cite{r26} and SIC \cite{r49,r50} can be employed. For the use of SIC in SUE, certain prerequisites must be met, including a minimum required SNR to decode the BS signal, ensuring channel symmetry, and availability of channel state information (CSI) at the SUE to facilitate correct decoding of information \cite{xiang2019cache}. On the other hand, we have considered the CSR setup in this paper; therefore, we can detect the BS signal using the ML detector. By subtracting it from the received signal, the SBD signal can be extracted \cite{r26}. This is possible because the decoding strategy for CSR treats the SBD signal as a multipath component rather than interference \cite{8907447}. Also, in SR system, the BS and SBDs work with each other in collaborative manner. This feature enables joint design of the BS and SBDs in a way that ensures any receiver can correctly decode their signals\cite{r25}. Ultimately, according to the above explanations, the BS signal does not create significant challenges for decoding the SBD information in this scheme.

Also, in a real network, SBDs are randomly distributed and each can be covered by a BS antenna. In this case, Each of these devices uses a unique carrier frequency that is transmitted by the nearest BS, causing the network to resemble a typical frequency division multiple access (FDMA) system. Therefor, there will be no interference in the network \cite{r46}. When multiple SBDs are covered by a BS, they may use the same carrier frequency for MTI sections to transmit information simultaneously. However, when a SUE is covered by multiple SBDs, there is a possibility of interference (${I_{0}}$) which is the worst-case scenario for creating interference in the SR network \cite{r25}. 

Several articles propose solutions to address this issue. For instance, article \cite{r40}, suggests utilizing mutual coding and decoding algorithms in SBDs to generate orthogonal chips that enable interference-free transmission among multiple SBDs. This creates a system that achieves multi-user symbiosis. Another article \cite{rn98} discusses the random distribution of SBDs in the network, which causes signals to reach the receiver at different power levels. To tackle this, a SIC process is employed to eliminate interference from signals with higher power levels than the desired SBD signal. Also, in \cite{r41} the SBD's information is backscattered to the SUE by multiplying the non-orthogonal random codes assigned to it, while \cite{turbocharging} introduces a novel orthogonal coding technique called $\mu$code, which offers the benefits of code division multiple access (CDMA), reduces interference, and enables concurrent transmissions. Therefore, according to the above explanation, by changing the transmit power of each SBD through either changing the power of the transmitted signal from the BS antenna or changing the reflection coefficients of each SBD, and ultimately using the SIC technique in SUE or assigning different orthogonal or non-orthogonal codes to each SBD, it is possible to reduce the interference ${I_{0}}$, caused by the concurrent transmission of their signals on the desired SBD signal in SUE, and avoid it.

Finally, in accordance with the above description and assuming that the beamforming vector of the BS has a unit power, the SNR for decoding \({s_{i,(j \in \upsilon_i )}}\)  in the SUE is:
\begin{equation}SNR_i^{UE} = \frac{{{p _i}{K}{{\left| {g_i^{}{\bf{h}}_i^H{{\bf{x}}_j}}\right|}^2}}}{{\sigma _{UE}^2}}\end{equation}

The proposed SR network uses SBDs that are passive and do not have active RF components. As a result, they do not emit radiation power and can only backscatter modulated signals. In this case, the only energy consuming component in the SBD is the microcontroller, that controls switches and impedances.
Therefore, the circuit energy consumption of the SBDs is equal to the energy required for backscattering the modulated signal, which can be calculated as ${p _i}{\tau _{j \in \upsilon_i}}$. It can be assumed that each SBD uses almost all of the harvested energy to backscatter their signals. Hence, the relation ${p _i}{\tau _{j \in \upsilon_i}} \approx \sum\limits_{j = 1,j \notin \upsilon_i }^J {{\varepsilon _{ij}}} $ is established, and Eq. (7) can be rewritten as follows:
 \begin{equation}
SNR_i^{UE} = \frac{{{{{K}\left| {g_i^{}{\bf{h}}_i^H{{\bf{x}}_j}} \right|}^2}\sum\limits_{j = 1,j \notin \upsilon_i }^J {{\varepsilon _{ij}}} }}{{{\tau _{j \in \upsilon_i}}\sigma _{UE}^2}}
\end{equation}

Therefore, the instantaneous achievable rate per unit bandwidth (SE) (bps/Hz) of the ${\text{SB}}{{\text{D}}_{\text{i}}}$ with considering the normalized bandwidth of the channel to 1$Hz$, is:
\begin{equation}
{R_i} = \frac{{\tau _{j \in \upsilon_i }}}{K}{\log _2}\left( {1+SNR_i^{UE}} \right)
\end{equation}

As we have used the CSR setup in this article, therefore, the spectrum growth phenomenon does not occur \cite{guo2019cognitive}, and by using the matched filter of the SUE, the BS and SBD signals can be well decoded. Therefore, there is no need for synchronization between SBDs and BS. It should be noted that SBDs are passive devices, and implementing traditional synchronization methods on them is also challenging. For easier access, the main notations used in this article have been introduced in Table I.
\begin{table}
    \centering
    \caption{Description of the main notaition of the paper.}
    \begin{tabular}{|c|c|}
        \Xhline{1.2pt} 
        \textbf{Notation} & \textbf{Description}\\
        \Xhline{1.2pt} 
        ${\tau _j}$ & Duration of $j$-th time slot , $ j = 1,2,...,J$ \\
        \hline
        ${{\rm{E}}_{\rm{T}}}$ & Energy consumption in the SR network \\
        \hline
        ${{\bf{x_j}}}$ & Ambient signal in ${\tau _j}$ with power ${p_{BS}}$ \\
        \hline
        ${s_{i,(j \in \upsilon_i) }}$ & Information signal of ${\text{SBD}_{\text{i}}}$ \\
        \hline
        ${{\bf{h}}_i}$ & Complex channel vector from BS to ${\text{SB}}{{\text{D}}_{\text{i}}}$\\
        \hline
         ${g_i}$ & Complex channel from ${\text{SB}}{{\text{D}}_{\text{i}}}$ to SUE\\
          \hline
        ${\varepsilon _{ij}}$ & Energy harvested by ${\text{SB}}{{\text{D}}_{\text{i}}}$(${\tau _j\notin \upsilon_i }$) \\
        \hline
        ${p _i}$&Power reflection coefficient by ${\text{SB}}{{\text{D}}_{\text{i}}}$\\
         \hline
        ${\eta _i}$ & Energy conversion efficiency by ${\text{SB}}{{\text{D}}_{\text{i}}}$\\
        \hline
        $y_{i,(j \in \upsilon_i)}^{UE}$ & Received signal at SUE, sent from BS, ${\text{SB}}{{\text{D}}_{\text{i}}}$\\
        \hline
        ${T}$ & Duration of time frame\\
        \hline
        $K$ & Number of BS symbol against each SBD symbol\\
        \hline
    \end{tabular}
    \label{table:example}
\end{table}

\section{ENERGY AND RATE OPTIMIZATION}
In this section, we present an optimization problem aimed at minimizing the total transmit energy of all BSs by jointly optimizing the transmit power of signal, time scheduling, and energy harvesting by SBDs. Additionally, the energy of the signal transmitted by each BS must satisfy the minimum information rate requirement of the corresponding SBD, denoted as \({C_i}\). To ensure that each SBD can transmit its information to the destination (SUE) at a minimum rate of \({C_i}\), we use the shannon channel capacity formula to specify the minimum channel capacity required between each SBD and the SUE, and we enforce this constraint for all SBDs.

According to the Eq.(3), Eq.(5) and Eq.(9),  the general optimization problem for the mentioned objectives and constraints is defined as follows:\begin{subequations}
\begin{align}
\mathop {{\rm{min }}}\limits_{\ {{{\bf{x}}_j},{\tau _j},{\varepsilon _{ij}}}} \,\,{{\rm{E}}_{\rm{T}}} = \sum\limits_{j = 1}^J {{\tau _j}{\bf{x}}_j^H{\bf{x}}_j^{}} \,\,\,\,\,\,\,\,\,\,\,\,\,\,\,\,\,\,\,\,\,\,\,\,\,\,\,\,\,\,\,\,\,\,\,\,\,\,\,\,\,\,\,\,\,\,\tag{10}\\
\,\,\,\,\,\,\,\,\,{\rm{s}}{\rm{.t}}{\rm{.}}\,\,\,\,\,\,\,\,\,{R_i} \ge {C_i}\,\,\,\,\,\,\,\,\,\,\,\,\,\,\,\,\,\,\,\,\,\,\,\,\,\,\,\,\,\,\,\,\,\,\,\,\,\,\,\,\,\,\,\,\,\,\,\,\,\,\,\,\,\, i \in \bf{\Psi} \label{eq:10a}\\
\,\,\,\,\,\,\,\,\,\,\,\,\,\,\,\,\,\,\,\,\,\,\,\,\,\,\,\,\,\,{\tau _j} \ge 0{\rm{ }} \,\,\,\,\,\,\,\,\,\,\,\,\,\,\,\,\,\,\,\,\,\,\,\,\,\,\,\,\,\,\,\,\,\,\,\,\,\,\,\,\,\,\,\,\,\,\,\,\,\,\,\,\,\,\,\,\,  j \in \bf{\Psi}  \label{eq:10b}\\
\,\,\,\,\,\,\,\,\,\,\,\,\,\,\,\,\,\,\,\,\,\,\,\,\,\,\,\,\,\sum\limits_{j = 1}^J {{\tau _j}}  \le T  {\rm{ }}\,\,\,\,\,\,\,\,\,\,\,\,\,\,\,\,\,\,\,\,\,\,\,\,\,\,\,\,\,\,\,\,\,\,\,\,\,\,\,\,\,\,\,\,\,\,j \in {\bf{\Psi}}  \label{eq:10c}\\
\,\,\,\,\,\,\,\,\,\,\,\,\,\,\,\,\,\,\,\,\,\,\,\,\,\,\,\,\,\,{\varepsilon _{ij}} \le {\eta _i}{\tau _{j \notin \upsilon_i }}{\bf{x}}_j^H{\bf{h}}_i^{}{\bf{h}}_i^H{{\bf{x}}_j}{\rm{    }} \,\,\,\,\,\,\,\,\,\,\, i,j \in \bf{\Psi}
\end{align}
\end{subequations}
where (10a) guarantees the minimum throughput requirements for the
${\text{SB}}{{\text{D}}_{\text{i}}}$ link, while equations (10b) and (10c) impose constraints on the total duration of all SBDs time slots within a time frame $T$, and (10d) restricts the energy harvested by the ${\text{SB}}{{\text{D}}_{\text{i}}}$ to its maximum specified value.

The problem of Eq.(10) is non-convex due to objective function and constraint (10d). However, constraint (a) is convex since it can be seen as a composition of perspective of $\log \left( {1 + \alpha F} \right),a > 0$ and an affine function of $F = {\varepsilon _{ij}}$.
To overcome the non-convexity of objective function, an auxiliary variable \(\gamma _j^{} = {\bf{x}}_j^H{\bf{x}}_j^{}\) is defined and as a result, another non-convex constraint appears. It can be relaxed by adding the following convex constraint:
\begin{equation}
{\gamma _j} \ge {\bf{x}}_j^H{\bf{x}}_j^{}\,\,\,,j \in  \bf{\Psi}
\end{equation}

So, the new problem is:\begin{subequations}
\begin{align}
\mathop {{\rm{min }}}\limits_{{{{\bf{x}}_j},{\tau _j},{\gamma _j},{\varepsilon _{ij}}}} {{\rm{E}}_{\rm{T}}} = \sum\limits_{j = 1}^J {{\tau _j}{\gamma _j}} \,\,\,\,\,\,\,\,\,\,\,\,\,\,\,\,\,\,\,\,\,\,\,\,\,\,\,\tag{12} \\
\,\,\,\,\,\,\,\,\,\,\,\,{\rm{s}}{\rm{.t}}  \,\,\,\,\,\,\,\,\,\, (10a),(10b),(10c),(10d)\\
  \,\,\,\,\,\,\,\,\,\,\,\,\,\,\,\,\,\,\,\,\,\,\,\,\,\,\,\,{\gamma _j} \ge {\bf{x}}_j^H{\bf{x}}_j^{}\,\,\,\,\,\,\,\,\,\,\,\,\,\,\,\,\,j \in  \bf{\Psi}
\end{align}
\end{subequations}

Although problem (12) remains non-convex, it contains a logarithmic function that is convex but not disciplined-convex. To solve disciplined-convex problems efficiently, we can use the primal-dual interior point method along with mature solver technologies. In subsections A, B, C, and D, we investigate the non-convex components of the general optimization problem and apply various techniques to transform it into a convex and discipline problem.

\subsection{Overcoming Non-Convex Objective Function}

To overcome the non-convexity of objective function we can rewrite it as follows:
 \begin{equation}
\sum\limits_{j = 1}^J {{\tau _j}\gamma _j^{}}  = \frac{1}{4}\left[ {\sum\limits_{j = 1}^J {{{\left( {{\tau _j} + \gamma _j^{}} \right)}^2}}  - \sum\limits_{j = 1}^J {{{\left( {{\tau _j} - \gamma _j^{}} \right)}^2}} } \right],j \in  \bf{\Psi}
\end{equation}

Eq.(13) is the difference between two convex functions. By definition ${f_1} \buildrel \Delta \over = {\left( {{\tau _j} - \gamma _j^{}} \right)^2}$, the second term  $ - {f_1}$ is a concave function and its upper bound can be calculated by a linear function. The upper bound of that relation performed by inner approximation \cite{r51} and sub-gradient method \cite{r52} around the feasible (initial) point \(\left( {{{\hat \tau }_j}^{},\hat \gamma _j^{}} \right)\) as follows:
 \begin{equation}
\begin{array}{l}
{f_1} \le {\left( {{{\hat \tau }_j} - \hat \gamma _j^{}} \right)^2} + {\nabla _{{\tau _j},\gamma _j^{}}}{f_1}\left( {{{\hat \tau }_j},\hat \gamma _j^{}} \right){\left[ {\left( {{\tau _j} - {{\hat \tau }_j}} \right),\left( {\gamma _j^{} - \hat \gamma _j^{}} \right)} \right]^T}\\
\,\,\,\,\,\,\,\, = {\left( {{{\hat \tau }_j} - \hat \gamma _j^{}} \right)^2} + 2\left( {{{\hat \tau }_j} - \hat \gamma _j^{}} \right)\left[ {\left( {{\tau _j} - {{\hat \tau }_j}} \right) - \left( {\gamma _j^{} - \hat \gamma _j^{}} \right)} \right]
\end{array}
\end{equation}

By replacing Eq. (14) in the second part of Eq. (13), we obtain an upper-bounded function that serves as a tangent line to the original function. This approximation function closely approximates the original function, particularly near the point of tangency, which corresponds to the point in Eq. (13) where the replacement was made. If we iteratively solve and converge Eq. (14), we can ensure that both Eq. (13) and Eq. (14) share the same local optimal point.

\subsection{Semidefinite Relaxation of Constraint (10d)}
By using the trace commutative property, we have the following equation:
  	\begin{equation}
{\bf{x}}_j^H{\bf{h}}_i^{}{\bf{h}}_i^H{{\bf{x}}_j} =Tr\left( {{\bf{x}}_j^H{\bf{h}}_i^{}{\bf{h}}_i^H{{\bf{x}}_j}} \right) =Tr\left( {{{\bf{x}}_j}{\bf{x}}_j^H{\bf{h}}_i^{}{\bf{h}}_i^H} \right)
\end{equation}

Now the constraint $\left(10d \right)$  can be reformulated by an auxiliary matrix  \({{\bf{X}}_j} = {{\bf{x}}_j}{\bf{x}}_j^H\) and \({{\bf{H}}_i} \buildrel \Delta \over = {\bf{h}}_i^{}{\bf{h}}_i^H\)  as follows:
\begin{equation}
{\varepsilon _{ij}} \le {\eta _i}{\tau _{j \notin \upsilon_i }}Tr\left( {{{\bf{X}}_j}{{\bf{H}}_i}} \right) \,\,\,\,\,\,\,\,\,\,\,\,\,\,\,\,\,\, i,j \in\bf{\Psi}
\end{equation}

 	So, since ${{\bf{X}}_j}$ is a positive semidefinite matrix, the constraint  $(16)$ can be replaced by: 
{\small\begin{subequations}
\begin{align}
{{{\eta _i}{\tau _{j \notin \upsilon_i }}Tr\left( {{{\bf{X}}_j}{{\bf{H}}_i}} \right)} \mathord{\left/
 {\vphantom {{{\eta _i}{\tau _{j \notin \upsilon_i }}Tr\left( {{{\bf{X}}_j}{{\bf{H}}_i}} \right)} {{\varepsilon _{ij}}}}} \right.
 \kern-\nulldelimiterspace} {{\varepsilon _{ij}}}} \ge 1\,\,\,\,\,\,\,\,\,\,\,\,\, i,j \in {\mathbf{\Psi }}\\
{{\bf{X}}_j} \succeq 0\,\,\,\,\,\,\,\,\,\,\,\,\,\,\,\,\,\,\,\,\,\,\,\,\,\,\,\,\,\,\,\,\,\,\,\,\,\,\,\,\,\,\,\,\,\,\,\,\,\,\,\,\,\,\,\,\,\,\,\,\,\,\,\,\,\,\,j \in {\mathbf{\Psi }}\\
{\rm{Rank}}\left( {{{\bf{X}}_j}} \right) = 1 \,\,\,\,\,\,\,\,\,\,\,\,\,\,\,\,\,\,\,\,\,\,\,\,\,\,\,\,\,\,\,\,\,\,\,\,\,\,\,\,\,\,\,\,\,\,\,j \in {\mathbf{\Psi }}
\end{align}
\end{subequations}}
where the notation $\succeq$ denotes that \({{\mathbf{X}}_j}\)  is a positive semidefinite. The problem is still non-convex due to the rank-one constraint given in  ${(17c)}$. By applying the semidefinite relaxation (SDR) technique, the rank-one constraint will be dropped, and the relaxed version of the main problem will be obtained. The relaxed problem will be a convex SDP problem that is solvable by interior-point methods. If the optimal solution \({{\mathbf{X}}_j}\) for the problem with the relaxed constraints (17) is rank-one, then it is also a solution for the original problem (12), which can be done by using randomization techniques \cite{r53}. In this state, we can easily extract a feasible \({{\mathbf{x}}_j}\)  from \({{\mathbf{X}}_j}\)  .

Moreover, we know ${\rm{Tr(}}{{\bf{X}}_j}{\rm{)}} \le \lambda _{\max }^{}({{\bf{X}}_j})$, where ${f_2} \buildrel \Delta \over = \lambda _{\max }^{}({{\bf{X}}_j})$  denotes the maximum eigenvalue of the ${{\bf{X}}_j}$ and it can be rewritten as the following difference between two convex functions \cite{r54}:
\begin{equation}
Tr\left( {{{\bf{X}}_j}} \right){\rm{ - }}\lambda _{\max }^{}\left( {{{\bf{X}}_j}} \right) \le 0
\end{equation}

To address the non-smooth nature of the rank one constraint, we employ the sub-gradient method. Therefore the sub-gradient of  $\lambda _{\max }^{}({{\bf{X}}_j})$  is:
{\small\begin{equation}
{\nabla _{{{\bf{X}}_j}}}{f_2}\left( {{{{\bf{\hat X}}}_j}} \right) = {{\bf{v}}_{\max }}\left( {{{\bf{X}}_j}} \right){\bf{v}}_{\max }^H\left( {{{\bf{X}}_j}} \right)
\end{equation}}
where \({{\bf{v}}_{\max }}\left( {{{\bf{X}}_j}} \right)\) is the maximum eigenvector corresponding to the maximum eigenvalue of ${{\bf{X}}_j}$. Therefore, the lower bound of linear approximation $\lambda _{\max }^{}({{\bf{X}}_j})$  at the point ${{\bf{\hat X}}_j}$  is:
{\small\begin{equation}
\begin{array}{l}
{f_2} \ge \lambda _{\max }^{}\left( {{{{\bf{\hat X}}}_j}} \right) + \left\langle {{\nabla _{{{\bf{X}}_j}}}{f_2}\left( {{{{\bf{\hat X}}}_j}} \right),\left( {{{\bf{X}}_j} - {{{\bf{\hat X}}}_j}} \right)} \right\rangle \\
\,\,\,\,\,\,\, = \lambda _{\max }^{}\left( {{{{\bf{\hat X}}}_j}} \right) + {\bf{v}}_{\max }^H\left( {{{{\bf{\hat X}}}_j}} \right){{\bf{v}}_{\max }}\left( {{{{\bf{\hat X}}}_j}} \right)\left( {{{\bf{X}}_j} - {{{\bf{\hat X}}}_j}} \right)
\end{array}
\end{equation}}

By substituting Eq.(20) in the Eq.(18) the following relationship is obtained: 
{\small\begin{equation}
Tr\left( {{{\bf{X}}_j}} \right) \le {\bf{v}}_{\max }^H\left( {{{{\bf{\hat X}}}_j}} \right){{\bf{X}}_j}{{\bf{v}}_{\max }}\left( {{{{\bf{\hat X}}}_j}} \right)
\end{equation}}

In order to solve the original Eq. (12), Eq. (21) is introduced as a convex subset of the original equation's solution space. To enforce this constraint, a penalty function is added to the objective function, where the coefficient  \(\ell \) determines the strength of the penalty. To ensure numerical stability, it is recommended to solve the resulting problem iteratively, gradually increasing the value of  \(\ell \) from small to large values. This approach avoids potential numerical issues and helps to converge to a feasible solution \cite{r55}.

Herein, the constraint ${(17a)}$ can be represented by the difference between two convex functions method and by defining the auxiliary variable \({\varepsilon _{ij}} \buildrel \Delta \over = \phi _{ij}^2\):
\begin{equation}
\phi _{ij}^2 \le {\eta _i}{\tau _{j \notin \upsilon_i }}Tr\left( {{{\bf{X}}_j}{{\bf{H}}_i}} \right)
\end{equation}

The above inequality is equivalent to:
{\small\begin{equation}
\phi _{ij}^2 + \frac{1}{4}{\left( {{\eta _i}Tr\left( {{{\bf{X}}_j}{{\bf{H}}_i}} \right) - {\tau _{j \notin \upsilon_i }}} \right)^2} \le \frac{1}{4}{\left( {{\eta _i}Tr\left( {{{\bf{X}}_j}{{\bf{H}}_i}} \right) + {\tau _{j \notin \upsilon_i }}} \right)^2}
\end{equation}}

	Since \(\,Tr\left( {{{\bf{X}}_j}{{\bf{H}}_i}} \right) + {\tau _{j \notin \upsilon_i }} \ge 0\), the above CQR relation can be written in the following form:
 \begin{equation}
\begin{array}{l}
{\left\| {\,\phi _{ij}^{},\frac{1}{2}\left( {{\eta _i}Tr\left( {{{\bf{X}}_j}{{\bf{H}}_i}} \right) - {\tau _{j \notin \upsilon_i }}} \right)\,} \right\|_2} \le \\
 \,\,\,\,\,\,\,\,\,\,\,\,\,\,\,\,\,\,\,\,\,\,\,\,\,\,\,\,\,\,\,\,\,\,\,\,\,\,\,\,\,\,\,\,\,\,\,\,\,\,\,\,\,\,\,\,\,\,\,\,\,\frac{1}{2}\left( {{\eta _i}Tr\left( {{{\bf{X}}_j}{{\bf{H}}_i}} \right) + {\tau _{j \notin \upsilon_i }}} \right)
\end{array}
\end{equation}	

Also, the constraint Eq.(12a) can be written as follows:
\begin{equation}
{\gamma _j} \ge Tr\left( {{{\bf{X}}_j}} \right)
\end{equation}

By changing the variable  \({\varepsilon _{ij}}\)=\(\phi _{ij}^2\), the constraint $(10a)$ will be non-convex versus the variables  \(\left( {\phi _{ij}^{},{\tau _{j \in \upsilon_i }}} \right)\) as it is not a perspective of function $\log \left( {1 + x} \right)$ anymore. In addition, any optimization problem that includes a $\rm{log}$  function, will turn the problem into a not disciplined convex problem and cannot be efficiently solved using modern SDP solvers, such as SeDuMi \cite{r56}. To overcome this problem, new solutions called sequential quadratic (SQ) and conic quadratic representation (CQR) methods are proposed that will be described in subsections C and D.

\subsection{Sequential Quadratic (SQ) }
An auxiliary  variable ${\theta _i}$ is defined to convert the constraint $a)$ of Eq.(10) to the two following inequalities:
{\small\begin{equation}
{a_1})\,\,{\theta _i} \le \frac{{{{{K}\left| {g_i^{}} \right|}^2}Tr\left( {{{\bf{X}}_j}{{\bf{H}}_i}} \right)}}{{\sigma _{UE}^2}}\sum\limits_{j = 1,j \notin \upsilon_i }^J {\phi _{ij}^2}   \,\,\,\,\ ,i,j \in  \bf{\Psi}
\end{equation}}{\small\begin{equation}
{a_2})\,\,{{K}C_i} -{ {\tau _{j \in \upsilon_i }}}{\log _2}\left( {1 + \frac{{{\theta _i}}}{{{\tau _{j \in \upsilon_i }}}}} \right) \le 0\,\,\,\,\,\,\,\,\,\,i \in \bf{\Psi}
\end{equation}}
The relation ${a_1})$ is not convex and should be using the difference of two convex functions method and linearization of the second part by the first-order approximation.  As mentioned in the previous sections, replacing the linear approximation around the initial point, in the second part of Eq.(26):
\begin{equation}
\begin{array}{l}
{\hat a_1})\,{\theta _i} -\frac{{K^2}\left|g_i\right|^4{Tr}({{{\bf{\hat{X}}}_j}\bf{H}}_i){\sum\limits_{j = 1,j \notin \upsilon_i }^J {\hat \phi _{ij}^2} }}{\sigma _{UE}^4}\times \\
\left(\begin{array}{l}
 {{Tr}({\bf{H}}_i)({{\bf{X}}_j}-{{\bf{\hat{X}}}_j}){\sum\limits_{j = 1,j \notin \upsilon_i }^J {\hat \phi _{ij}^2} }}+\\
{2\mathrm{Tr}({{{\bf{\hat {X}}}_j}\bf{H}}_i)({\phi _{ij}}-{\hat \phi _{ij}}){\sum\limits_{j = 1,j \notin \upsilon_i }^J {\hat \phi _{ij}} }}
\end{array}\right) \le 0
\end{array}
\end{equation}	
The constraint ${a_2}$ appears to be convex, but the presence of the logarithm function makes it non-disciplined. To address this, one approach is to find an upper bound approximation using a quadratic form and apply inner approximation techniques. A theorem provides an iterative approach that is similar to inner approximation and can converge to a local optimum for smooth optimization problems.

\textbf{Theorem}: the function ${f_3}:D \to \mathbb{R}$, \(\forall x,\hat x \in D\) , is convex if a second derivative exists at each point in domain $D$ and  \({\nabla ^2}f_3\left( {\hat x} \right) \ge 0\), hence, the second-order approximation is \cite{r57}:
 \begin{equation}
\begin{array}{l}
{f_3}\left( x \right) = {f_3}\left( {\hat x} \right) + \nabla {f_3}\left( {\hat x} \right){\left( {x - \hat x} \right)^T}\times \\
 \,\,\,\,\,\,\,\,\,\,\,\,\,\,\,\,\,\,\,\,\,\,\,\,\,\,\,\,\,\,\,\,\,\,\,\,\,\,\,\,\,\,\,\,\,\,\,\,\,\,\,\,\,\,\,\,\,\,\,\,\,\left( {x - \hat x} \right){\nabla ^2}{f_3}\left( {\hat x} \right){\left( {x - \hat x} \right)^T}
\end{array}
\end{equation}	
According to the above relation, the matrix \({{\bf{H}}_{\bf{s}}}\) (called an upper bound Hessian matrix) must satisfy the following relation:
{\small\begin{equation}
{\nabla ^2}{f_3}\left( {\hat x} \right) \le {{\bf{H}}_{\bf{s}}}
\end{equation}}
If the above relation is established, Eq.(29) can be converted as follows:
{\small\begin{equation}
\begin{array}{l}
{f_3}\left( x \right) \le {f_3}\left( {\hat x} \right) + \nabla {f_3}\left( {\hat x} \right){\left( {x - \hat x} \right)^T} + \left( {x - \hat x} \right){{\bf{H}}_{\bf{s}}}{\left( {x - \hat x} \right)^T}
\end{array}
\end{equation}}

where the function $f_3$ is defined as the left-hand side of the constraint \({a_2})\). The value of  ${\nabla ^2}{f_3}\left( {\hat x} \right)$ can be obtained as follows:
\begin{equation}
{\nabla ^2}{f_3}\left( {{\theta _i},{\tau _{j \in  \upsilon_i }}} \right)\, = \left[ {\begin{array}{*{20}{c}}
{\frac{{{\tau _{j \in  \upsilon_i }}}}{{{{\left( {{\tau _{j \in  \upsilon_i }} + {\theta _i}} \right)}^2}}}}&{ - \frac{{{\theta _i}}}{{{{\left( {{\tau _{j \in \upsilon_i }} + {\theta _i}} \right)}^2}}}}\\
{ - \frac{{{\theta _i}}}{{{{\left( {{\tau _{j \in  \upsilon_i }} + {\theta _i}} \right)}^2}}}}&{\frac{{\theta _i^2}}{{{\tau _{j \in  \upsilon_i }}{{\left( {{\tau _{j \in  \upsilon_i }} + {\theta _i}} \right)}^2}}}}
\end{array}} \right]
\end{equation}

Matrix \(\nabla _{}^2{f_3}\left( {{\theta _i},{\tau _{j \in \upsilon_i }}} \right)\) for ${\tau _{j \in \upsilon_i }},{\theta _i} \ge 0$  does not have an upper bound. Therefore, we consider the below feasible set to bound it ($\beta $ is a fixed number):
\begin{equation}
{D_i} = \left\{ {\left. {\left( {{\theta _i},{\tau _{j \in \upsilon_i }}} \right)} \right|{\theta _i} \ge 0,0 \le \frac{{{{\hat \tau }_{j \in \upsilon_i }}}}{\beta } \le {\tau _{j \in \upsilon_i }} \le \infty } \right\}
\end{equation}

This new domain, which is an implicit constraint, is added to the constraints of the main problem. Now to estimate an upper bound of Eq.(32), the following matrix is replaced instead of it ( \emph{proof}: Please refer to Appendix):
\begin{equation}
\nabla _{}^2{f_3}\left( {{\theta _i},{\tau _{j \in \upsilon_i }}} \right) \le \left[ {\begin{array}{*{20}{c}}
{\frac{{9\beta }}{{8{{\hat \tau }_{j \in \upsilon_i }}}}}&{ - \frac{\beta }{{8{{\hat \tau }_{j \in \upsilon_i }}}}}\\
{ - \frac{\beta }{{8{{\hat \tau }_{j \in \upsilon_i }}}}}&{\frac{{9\beta }}{{8{{\hat \tau }_{j \in \upsilon_i }}}}}
\end{array}} \right] \le {{\bf{H}}_{\bf{s}}}
\end{equation}

Also, the gradient of Eq.(27) is as follows:
{\small\begin{equation}
\nabla {f_3}\left( {{\theta _i},{\tau _{j \in \upsilon_i }}} \right)\, = \left[{\begin{array}{*{20}{c}}
{-\frac{{{\tau _{j \in \upsilon_i}}}}{{{\tau _{j \in \upsilon_i }} + {\theta _i}}}},{ - \log \left( {1 + \frac{{{\theta _i}}}{{{\tau _{j \in \upsilon_i }}}}} \right) + \frac{{{\theta _i}}}{{{\tau _{j \in \upsilon_i }} + {\theta _i}}}}
\end{array}} \right]
\end{equation}}

Therefore, according to Eqs.(27), (29), (34) and (35), the ${a_2)}$ constraint is converted to:
{\small\begin{equation}
{\hat a_2})\left( {\begin{array}{*{20}{l}}
{K{C_i} - {{\hat \tau }_{j \in {\upsilon _i}}}{{\log }_2}\left( {1 + \frac{{{{\hat \theta }_i}}}{{{{\hat \tau }_{j \in {\upsilon _i}}}}}} \right) + }\\
{\nabla {f_3}\left( {{{\hat \theta }_i},{{\hat \tau }_{j \in {\upsilon _i}}}} \right){{\left[ {{\theta _i} - {{\hat \theta }_i},{\tau _{j \in {\upsilon _i}}} - {{\hat \tau }_{j \in {\upsilon _i}}}} \right]}^T} + }\\
{\left[ {{\theta _i} - {{\hat \theta }_i},{\tau _{j \in {\upsilon _i}}} - {{\hat \tau }_{j \in {\upsilon _i}}}} \right]{{\bf{H}}_{\bf{s}}}{{\left[ {{\theta _i} - {{\hat \theta }_i},{\tau _{j \in {\upsilon _i}}} - {{\hat \tau }_{j \in {\upsilon _i}}}} \right]}^T}}
\end{array}} \right) \le 0
\end{equation}}

Finally, according to Eqs.(10), (12), (13), (14), (17), (21), (24), (25), (28), (33) and (36), the final convex optimization for the SQ approach is obtained as follows:
{\small\begin{equation}
\begin{array}{l}
\mathop {{\rm{min }}}\limits_{\scriptstyle\tau _j^{},{{\bf{X}}_j},\gamma _j^{}\hfill\atop
\scriptstyle\,\,\,\,\phi _{ij}^{},{\theta _i}\hfill} \overbrace {\sum\limits_{j = 1}^J {\left( {\begin{array}{*{20}{l}}
{\frac{1}{4}{{\left( {{\tau _j} + \gamma _j^{}} \right)}^2} - \frac{1}{4}{{\left( {{{\hat \tau }_j}^{} - \hat \gamma _j^{}} \right)}^2} - }\\
{\frac{1}{2}\left( {{{\hat \tau }_j}^{} - \hat \gamma _j^{}} \right)\left( {\left( {{\tau _j} - \gamma _j^{}} \right) - \left( {{{\hat \tau }_j} - \hat \gamma _j^{}} \right)} \right) + }\\
{\ell \left( {Tr\left( {{{\bf{X}}_j}} \right) - {{\bf{v}}_{{{\max }^H}}}\left( {{{{\bf{\hat X}}}_j}} \right){{\bf{X}}_j}{{\bf{v}}_{\max }}\left( {{{{\bf{\hat X}}}_j}} \right)} \right)}
\end{array}} \right)} }^{{{\rm{E}}_{\rm{T}}} = } \\
\,\,\,\,\,\,\,{\rm{s}}{\rm{.t}}{\rm{.}}\,\,\,\,\,\,\,\,\,\,\,\,(10b),(10c),\left( {17b} \right),(24),(25),(28),(33),(36)
\end{array}
\end{equation}}

Now this problem is discipline and convex. So, the optimal value of variables and objective function for the SQ approach is obtained by the algorithm 1. 
\begin{algorithm}
\caption{Sequential Quadratic (SQ) }\label{alg:alg1}
\begin{algorithmic}
\STATE 
\STATE 1-	Initialize ${{\bf{\hat X}}_j},{\hat \phi _{ij}},{\hat \tau _j},{\hat \gamma _j},{\hat \theta _i},\beta ,{K},a$ in the feasible set
\STATE 2-  Choose $\varepsilon  \ge 0,\eta_i \ge 0$
\STATE 3- \textbf{While} $counter < counte{r_{\max }}$
\STATE 4- \hspace{0.1cm} Solve (37) to obtain the solution variables \\ \,\,\,\,\,\,\,\,\,\, ${{\bf{X}}_j},{\phi _{ij}},{\tau _j},{\gamma _j},{\theta _i}$
\STATE 5- \hspace{0.1cm} \textbf{IF} $\left\{{\small\begin{array}{l}
\left\| {{{\bf{X}}_j} - {{{\bf{\hat X}}}_j}} \right\|,\left\| {{\phi _{ij}} - {{\hat \phi }_{ij}}} \right\|,\left\| {{\tau _j} - {{\hat \tau }_j}} \right\|,\\
\,\,\,\,\,\,\,\,\,\left\| {{\gamma _j} - {{\hat \gamma }_j}} \right\|,\left\| {{\theta _i} - {{\hat \theta }_i}} \right\|
\end{array}} \right\}> \varepsilon $ 
\STATE 6-  \hspace{0.6cm}\({{\bf{X}}_j} \to {{\bf{\hat X}}_j},{\phi _{ij}} \to {\hat \phi _{ij}},{\tau _j} \to \hat \tau ,{\gamma _j} \to {\hat \gamma _j},{\theta _i} \to {\hat \theta _i}\)
\STATE 7- \hspace{0.5cm} Go to step 3
\STATE 8- \hspace{0.1cm} \textbf{Else}
\STATE 9- \hspace{0.5cm}  Check the the Rank one constraint Eq.(18) to be \\ \,\,\,\,\,\,\,\,\,\,\,\,\,\,\,\, satisfied by \(\frac{{Tr}\left( {{\mathbf{X}}_{j}} \right)\text{-}\lambda _{\max }^{{}}\left( {{\mathbf{X}}_{j}} \right)}{{Tr}\left( {{\mathbf{X}}_{j}} \right)}\le \varepsilon\)
\STATE 10- \hspace{0.3cm} $counter \to counter + 1$
\STATE 11- \hspace{0.5cm}\textbf{IF}  the constraint is not satisfied in Step 9 
\STATE 12- \hspace{0.7cm} then set $\alpha \ell  \to \ell $ and go to Step 3
\STATE 13- \hspace{0.5cm}\textbf{Else } 
\STATE 14- \hspace{0.6cm}$\left( {{{\bf{X}}_{opt}},{\phi _{opt}},{\tau _{opt}},{\gamma _{opt}},{\theta _{opt}}} \right) = \left( {{{\bf{X}}_j},{\phi _{ij}},{\tau _j},{\gamma _j},{\theta _i}} \right)$
\STATE 15- \hspace{0.5cm}\textbf{End IF}
\STATE 16- \hspace{0.2cm}\textbf{End IF}
\STATE 17- \hspace{0 cm}\textbf{End While}
\end{algorithmic}
\label{alg1}
\end{algorithm}
\subsection{Conic Quadratic Representation (CQR)}
An auxiliary variable \({z_i}\) is introduced to convert the constraint $a)$ Eq.(10) to the two following inequalities:
{\small\begin{equation}
{a_1}){z_i} \ge \frac{{{{K}C_i}}}{{{\tau _{j \in \upsilon_i}}}}\,\,\,\,\,\,,i \in {\bf{\Psi }}
\end{equation}}{\small\begin{equation}
{a_2}){\log _2}\left( {1 + \frac{{{{{K}\left| {g_i^{}} \right|}^2}{Tr\left( {{{\bf{X}}_j}{{\bf{H}}_i}} \right)}}}{{{\tau _{j \in \upsilon_i }}\sigma _{UE}^2}}\sum\limits_{j = 1,j \notin \upsilon_i }^J {\phi _{ij}^2} } \right) \ge {z_i}^{},i \in {\bf{\Psi }}
\end{equation}}
To incorporate the mentioned constraints into the main optimization problem, it is necessary to ensure that these equations are both convex and disciplined. Therefore, we need to make specific modifications to meet these requirements. (consider \({z_i} \ge 0,i \in {\bf{\Psi }}\)): 
\begin{equation}
{z_i} \ge \frac{{{K}{C_i}}}{{{\tau _{j \in \upsilon_i }}}} \to {\left( {{z_i} + {\tau _{j \in \upsilon_i }}} \right)^2} \ge 4{{K}C_i} + {\left( {{z_i} - {\tau _{j \in \upsilon_i }}} \right)^2}
\end{equation}

Since \({z_i},{\tau _{j \in \upsilon_i }} \ge 0\) , the above CQR relation can be written as follows:
{\small\begin{equation}
\left( {{z_i} + {\tau _{j \in \upsilon_i }}} \right) \ge {\left\| {\left[ {2\sqrt {{{K}C_i}} ,\left( {{z_i} - {\tau _{j \in \upsilon_i }}} \right)} \right]} \right\|_2}  
\end{equation}}

To simplify the inequality Eq.(39), the new auxiliary variable \({\Xi _i}\) , \(i \in {\bf{\Psi }}\) is defined:
{\small\begin{equation}
{\Xi _i} \le \frac{{{{{K}\left| {g_i^{}} \right|}^2}{Tr\left( {{{\bf{X}}_j}{{\bf{H}}_i}} \right)}\sum\limits_{j = 1,j \notin \upsilon_i }^J {\phi _{ij}^2} }}{{{\tau _{j \in \upsilon_i}}\sigma _{UE}^2}}\,\,\,\,\,\,,\,\,\,\,\,i \in {\bf{\Psi }}
\end{equation}}

Therefore
{\small\begin{equation}
1 + {\Xi _i} \ge {e^{{z_i}}},\,\,i \in {\bf{\Psi }}
\end{equation}}

The constraint of Eq.(43) is non-convex. In general, to solve the above relation, it is needed to approximate it with the following lemma.

\textbf{Lemma}: If a set of auxiliary variables  \({\zeta _{q,i}},q \in \left\{ {1,...,M + 4} \right\},i \in {\bf{\Psi }} \) satisfies the following inequalities, then we can use the CQR method to approximate the equivalence of \(1 + {\Xi _i} \ge {e^{{z_i}}},\,\,i \in {\bf{\Psi }}\), which can be represented as the following set of linear and conic inequalities.
{\small\begin{subequations}
\begin{flalign*}
&1 + {\Xi _i} \ge {\zeta _{M + 4,i}},\tag{44a}\\
&1 + {\zeta _{1,i}} \ge {\left\| {\left[ {\begin{array}{*{20}{c}}
{1 - {\zeta _{1,i}}}&{2 + {2^{1 - M}}{z_i}}
\end{array}} \right]} \right\|_2},\tag{44b}\\
&1 + {\zeta _{2,i}} \ge {\left\| {\left[ {\begin{array}{*{20}{c}}
{1 - {\zeta _{2,i}}}&{{5 \mathord{\left/
 {\vphantom {5 3}} \right.
 \kern-\nulldelimiterspace} 3} + {2^{ - M}}{z_i}}
\end{array}}\right]} \right\|_2},\tag{44c}\\
&1 + {\zeta _{3,i}} \ge {\left\| {\left[ {\begin{array}{*{20}{c}}
{1 - {\zeta _{3,i}}}&{2{\zeta _{1,i}}}
\end{array}} \right]} \right\|_2},\tag{44d}\\
&{\zeta _{4,i}} \ge {\zeta _{2,i}} + {{{\zeta _{3,i}}} \mathord{\left/
 {\vphantom {{{\zeta _{3,i}}} {24}}} \right.
 \kern-\nulldelimiterspace} {24}} + {{19} \mathord{\left/
 {\vphantom {{19} {72}}} \right.
 \kern-\nulldelimiterspace} {72}},\tag{44e}\\
&1 + {\zeta _{q,i}} \ge {\left\| {\left[ {\begin{array}{*{20}{c}}
{1 - {\zeta _{q,i}}}&{2{\zeta _{q - 1,i}}}
\end{array}} \right]} \right\|_2}, q \in \left\{ {5,..., M + 4} \right\} \tag{44f}
\end{flalign*}
\end{subequations}}

The accuracy of the approximation can be improved by increasing the value of $M$. The optimal value of $M$ depends on finding a balance between achieving convergence and minimizing computational complexity. This value of $M$ is referred to as the approximation coefficient.

\emph{Proof}: Please refer to \cite{r6,r58}

To establish a relation between the variables \(\phi _{ij}^2\) and \({\Xi _i}\) the linear approximation for the right-hand side of Eq.(42) should be written similarly to previous sections. By placing it in Eq.(42), the new constraint is obtained as follows and should be added to other constraints of the main problem Eq.(12):
\begin{equation}
\begin{array}{l}
{\Xi _i}-\frac{{K^2}\left|g_i\right|^4\mathrm{Tr}({{{\bf{\hat{X}}}_j}\bf{H}}_i){\sum\limits_{j = 1,j \notin \upsilon_i }^J {\hat \phi _{ij}^2} }}{\hat \tau _{j \in \upsilon_i }\sigma _{UE}^4}\times\\
\left(\begin{array}{l}
\frac{\mathrm{Tr}({\bf{H}}_i)({{\bf{X}}_j}-{{\bf{\hat{X}}}_j}){\sum\limits_{j = 1,j \notin \upsilon_i }^J {\hat \phi _{ij}^2} }}{\hat \tau _{j \in \upsilon_i }}
-\\
\frac{\mathrm{Tr}({{{\bf{\hat {X}}}_j}\bf{H}}_i)(\tau _{j \in \upsilon_i }-\hat \tau _{j \in \upsilon_i }){\sum\limits_{j = 1,j \notin \upsilon_i }^J {\hat \phi _{ij}^2} }}{\hat \tau _{j \in \upsilon_i }^2}+\\
\frac{2\mathrm{Tr}({{{\bf{\hat {X}}}_j}\bf{H}}_i)({\phi _{ij}}-{\hat \phi _{ij}}){\sum\limits_{j = 1,j \notin \upsilon_i }^J {\hat \phi _{ij}} }}{\hat \tau _{j \in \upsilon_i }}
\end{array}\right) \le 0
\end{array}
\end{equation}	
Finally, according to Eqs, (10), (12), (13), (14), (17), (21), (24), (25), (41), (44) and (45), the final convex optimization for the CQR approach is obtained as follows:
{\small\begin{equation}
\begin{array}{l}
\mathop {{\rm{min}}}\limits_{\begin{array}{*{20}{c}}
{\scriptstyle\tau _j^{},{{\bf{X}}_j},\hfill\atop
\scriptstyle\gamma _j^{},{z_i},\hfill}\\
{\scriptstyle{\Xi _i},\phi _{ij}^{},\hfill\atop
\scriptstyle{\zeta _{q,i}}\hfill}
\end{array}}  \overbrace {\sum\limits_{j = 1}^J {\left( {\begin{array}{*{20}{l}}
{\frac{1}{4}{{\left( {{\tau _j} + \gamma _j^{}} \right)}^2} - \frac{1}{4}{{\left( {{{\hat \tau }_j}^{} - \hat \gamma _j^{}} \right)}^2} - }\\
{\frac{1}{2}\left( {{{\hat \tau }_j}^{} - \hat \gamma _j^{}} \right)\left( {\left( {{\tau _j} - \gamma _j^{}} \right) - \left( {{{\hat \tau }_j} - \hat \gamma _j^{}} \right)} \right) + }\\
{\ell \left( {Tr\left( {{{\bf{X}}_j}} \right) - {{\bf{v}}_{{{\max }^H}}}\left( {{{{\bf{\hat X}}}_j}} \right){{\bf{X}}_j}{{\bf{v}}_{\max }}\left( {{{{\bf{\hat X}}}_j}} \right)} \right)}
\end{array}} \right)} }^{{{\rm{E}}_{\rm{T}}} = }  \\
\,\,\,\,\,\,\,\,\,{\rm{s}}{\rm{.t}}{\rm{.}}\,\,\,\,\,\,\,\,\,\,\,\,(10b),(10c),\left( {17b} \right),(24),(25),(41),(44),\left( {45} \right)
\end{array}
\end{equation}}
\begin{algorithm}
\caption{Conic Quadratic Representation (CQR) }
\begin{algorithmic}
\STATE 
\STATE 1-	Initialize ${{\bf{\hat X}}_j},{\hat \phi _{ij}},{\hat \tau _j},{\hat \gamma _j},\alpha ,{K},M$ in the feasible set
\STATE 2-  Choose $\varepsilon  \ge 0,\eta_i  \ge 0$
\STATE 3- \textbf{While} $counter < counte{r_{\max }}$
\STATE 4- \hspace{0.1cm} Solve (46) to obtain the solution variables \\ \,\,\,\,\,\,\,\,\,\,${{\bf{X}}_j},{\phi _{ij}},{\tau _j},{\gamma _j},{\Xi _i},{\xi _{M + 4,i}},{z_i}$
\STATE 5- \hspace{0.1cm} \textbf{IF} {\small $\left\{ \begin{array}{l}
\left\| {{{\bf{X}}_j} - {{{\bf{\hat X}}}_j}} \right\|,\left\| {{\phi _{ij}} - {{\hat \phi }_{ij}}} \right\|,\\
\left\| {{\tau _j} - {{\hat \tau }_j}} \right\|,\left\| {{\gamma _j} - {{\hat \gamma }_j}} \right\|
\end{array} \right\} > \varepsilon $}
\STATE 6-  \hspace{0.6cm}\({{\bf{X}}_j} \to {{\bf{\hat X}}_j},{\phi _{ij}} \to {\hat \phi _{ij}},{\tau _j} \to \hat \tau ,{\gamma _j} \to {\hat \gamma _j}\)
\STATE 7- \hspace{0.5cm} Go to step 3
\STATE 8- \hspace{0.1cm} \textbf{Else}
\STATE 9- \hspace{0.5cm} Check the the Rank one constraint Eq.(18) to be \\ \,\,\,\,\,\,\,\,\,\,\,\,\,\,\,\, satisfied by \(\frac{{Tr}\left( {{\mathbf{X}}_{j}} \right)\text{-}\lambda _{\max }^{{}}\left( {{\mathbf{X}}_{j}} \right)}{{Tr}\left( {{\mathbf{X}}_{j}} \right)}\le \varepsilon\)
\STATE 10- \hspace{0.3cm} $counter \to counter + 1$
\STATE 11- \hspace{0.5cm}\textbf{IF} the constraint is not satisfied in Step 9  
\STATE 12- \hspace{0.7cm} then set $\alpha \ell  \to \ell $ and go to Step 3
\STATE 13- \hspace{0.5cm}\textbf{Else } 
\STATE 14- \hspace{0.7cm}$\left( {{{\bf{X}}_{opt}},{\phi _{opt}},{\tau _{opt}},{\gamma _{opt}}} \right) = \left( {{{\bf{X}}_j},{\phi _{ij}},{\tau _j},{\gamma _j}} \right)$
\STATE 15- \hspace{0.5cm}\textbf{End IF}
\STATE 16- \hspace{0.2cm}\textbf{End IF}
\STATE 17- \hspace{0 cm}\textbf{End While}
\end{algorithmic}
\label{alg1}
\end{algorithm}

Now this problem is discipline and convex. So, the optimal value of variables and objective function for the CQR approach is obtained by algorithm 2. The advantage of the CQR method is to obtain the optimal solution by increasing the value of the coefficient of approximation ($M$). At the end of this calculation and after obtaining ${{\bf{X}}_j}$, the variable ${{\bf{x}}_j}$ is obtained according to Equation ${{\bf{X}}_j} = {{\bf{x}}_j}{\bf{x}}_j^H$.
Finally, the proposed methods can be solved iteratively by starting from the initial feasible point and performing successive updates until convergence is achieved.

\section{COMPUTATIONAL COMPLEXITY}
In this section, we present complexity bounds in terms of real arithmetic operations before an $\bm{\varepsilon}$-solution is obtained for the proposed optimization problems using the interior-point method \cite{r59}. Interior point methods were extended from linear optimization to semi-definite optimization and the polynomial complexity of the algorithm can be obtained theoretically \cite{r60}.
As you observed, the CQR and SQ methods were implemented as Second-order cone program (SOCP) and SDP models, respectively. The computational complexities of these two methods are shown in Table II.
\begin{table}
    \caption{The computational complexity of the SQ and CQR methods.}
\begin{tabular}{|c|}
  \Xhline{1.2pt} 
  \textbf{SQ method (Algorithm 1)} \\
  \Xhline{1.2pt} 
  $\begin{aligned}
    &\mathcal{O}(1)\sqrt{1 + \sum_{v = 1}^V \kappa_v}\left(B^3 + B^2\sum_{v = 1}^V \kappa_v^2 + B\sum_{v = 1}^V \kappa_v^3\right)\ln\left(\frac{\text{size}(SQ)}{\varepsilon}\right)\\
    & \,\,\,\,\,\,\,\,\,\,\,\,\,\,\,\,\text{size}(SQ) = \left( {\left( {B + 1} \right)\sum\limits_{v = 1}^V {\frac{{{\kappa _v}\left( {{\kappa _v} + 1} \right)}}{2}} } \right) + V + B + 3\\
    &\,\,\,\,\,\,\,\,\,\,\,\,\,\,\,\,\,\,\,\,\,\,\,\,\,\,\kappa_v = I\left(2N^2 + I + 1.5\right),\quad B = 5,\quad V = 8
  \end{aligned}$ \\
  \Xhline{1.2pt} 
  \textbf{CQR method (Algorithm 2)} \\
  \Xhline{1.2pt} 
  $\begin{aligned}
    &\mathcal{O}(1)\sqrt{V + 1}B\left(B^2 + V + \sum_{v = 0}^V \kappa_v^2\right)\ln\left(\frac{\text{size}(CQR)}{\varepsilon}\right)\\
    &\text{size}(CQR) = \left(V + \sum_{v = 1}^V \kappa_v\right)\left(B + 1\right) + V + B + 3\\
    &\kappa_v = I\left(3N^2 + 0.5I^2 + M + 7\right),\quad B = 7,\quad V = 13
  \end{aligned}$ \\
  \hline
\end{tabular}
\end{table}

In Table II, $Ln\left( {{{size\left( . \right)} \mathord{\left/ {\vphantom {{size\left( p \right)} \varepsilon }} \right. \kern-\nulldelimiterspace} \varepsilon }} \right)$ represents the number of accuracy digits in an $\varepsilon$-solution, ${size\left( . \right)}$ is the dimension of the total data of the QR or SQ problems, \(\kappa_{v} \) is the dimensions of the $v$-th constraint, $B$ and $V$ is the number of variables and a number of constraints of the optimization problem respectively.
To better understand the complexity of relationships, we have calculated these values and presented them in Table II. Additionally, the corresponding figure has been provided in the simulation section, which uses the exact results of these calculations.
\section{SIMULATION RESULTS}
According to the system model, SBDs are randomly distributed in the network and located near the SUE. Each IoT device can harvest energy and send its information to the intended SUE, as shown in Fig. 1. In all simulations, we consider $T = 10$, $K=100$, $\eta  = 0.8$ and \(\sigma _i^2=\sigma _{UE}^2 =  - 114\,\,dBm\). Also, we assume the carrier frequency is 2 GHz, the channel bandwidth is 400 kHz, the pathloss exponents are 3, and the BS antenna gain equals 5 dB. All simulation results were obtained by averaging over 1000 randomly generated channels, and ten different initialization points in the convex feasible set were considered to ensure the stability of the problems. The distance between the BS and SUE is 200 meters, and the maximum distance of SBDs from SUE is 100 meters, which are distributed randomly between the BS and SUE. All simulations were performed using a laptop with a Core i7 processor and 8 GB RAM.
\subsection{Comparison of Proposed Methods}
This paper utilizes two mathematical methods, CQR and SQ, to solve the optimization problem. In the CQR method, the first step involves obtaining the appropriate approximation coefficient. To do so, we plot the diagram of the minimum energy consumption by the BS versus the SBDs throughput requirement, as shown in Fig. 3, assuming approximation coefficients of $M = 1, 2, ..., 6$ in the CQR method.
\begin{figure}
\begin{center}
\includegraphics[height=5.5cm, width=13cm]{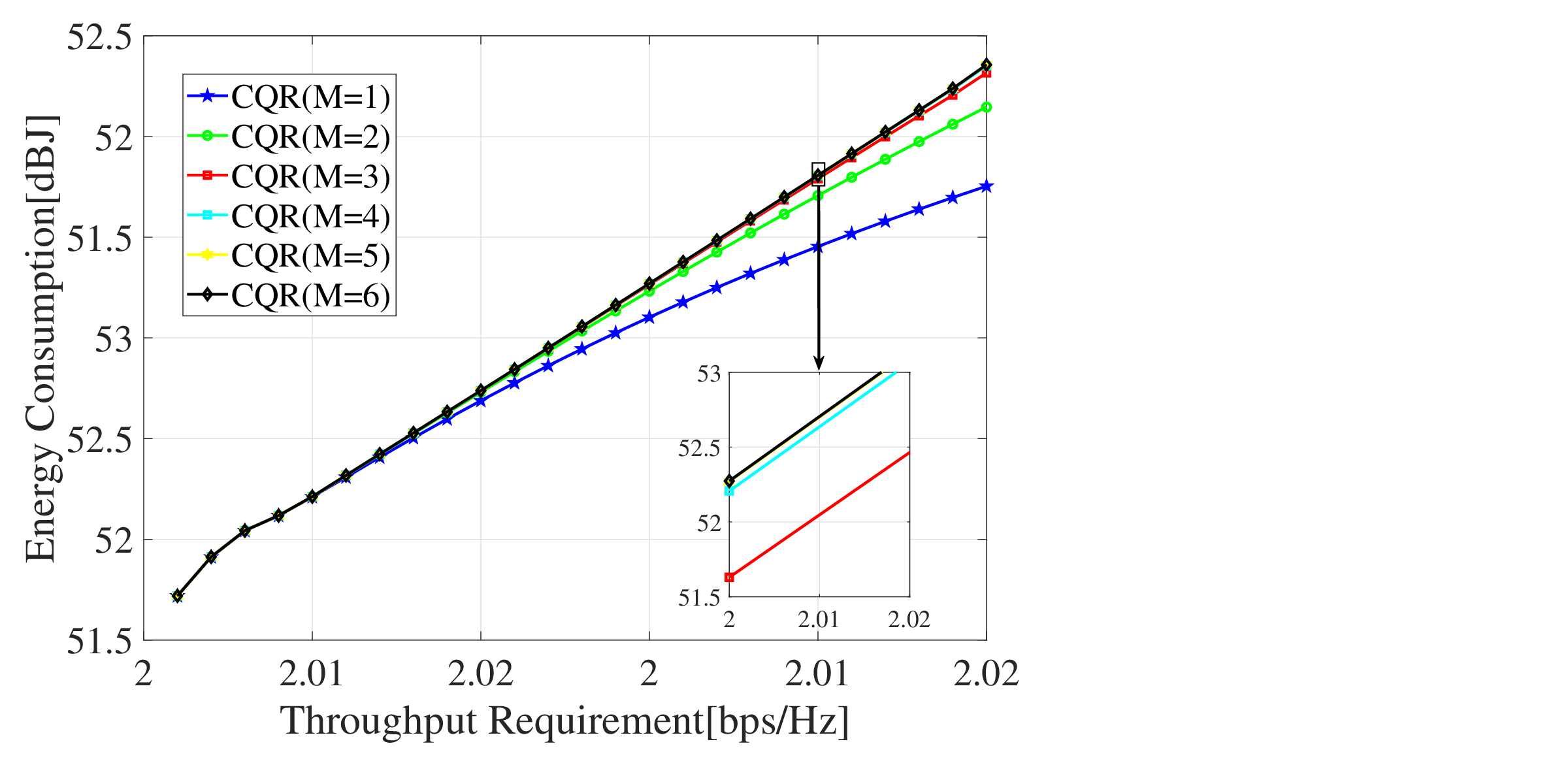}
\caption{EC versus SBDs throughput requirement in CQR method for \(M = 1,2,...,6\), \(I = 6\)}
\end{center}
\end{figure}
As shown in Fig. 3, the total energy consumption (EC) in the network transferred by the BS increases as the minimum data transmission rate in SBDs increases, as expected. To find the best approximation coefficient for this problem, we need to identify the point where the shapes converge. Based on Fig. 3, this event occurs for \(M \ge 4\), and therefore, the appropriate values can be obtained. By considering the computational complexity, as shown in Fig. 6, the optimal value is determined to be $M = 4$.

To determine which mathematical solution is more accurate in finding optimal points, we compare two methods: CQR with M = 4 and SQ. In this case, we change the basic parameters of the network design, including the number of IoT devices in the network. Fig. 4 shows the minimum energy transfer versus SBDs throughput requirement for these two methods, where the number of SBDs varies from 2 to 4  ($I \in \left\{ {2,3,4} \right\}$).

For example, in the special case of $I = 3$, the CQR method is more effective than the SQ method in minimizing the total EC in the network while still meeting the SBDs' throughput requirements. At a rate of 0.9 bps/Hz, the difference in EC between two cases of 2 and 4 SBDs is 22.71dB for the SQ method and 6.77dB for the CQR method. As the number of SBDs increases, the CQR method is more stable and has better EC compared to the SQ method.

Another significant parameter to consider is the number of energy output antennas on the BS. Fig. 5 illustrates this parameter for BSs with different numbers of antennas, including single antenna, $2 \times 2$, and $3 \times 3$ antennas configurations.
\begin{figure}
\begin{center}
\includegraphics[height=5.5cm, width=13cm]{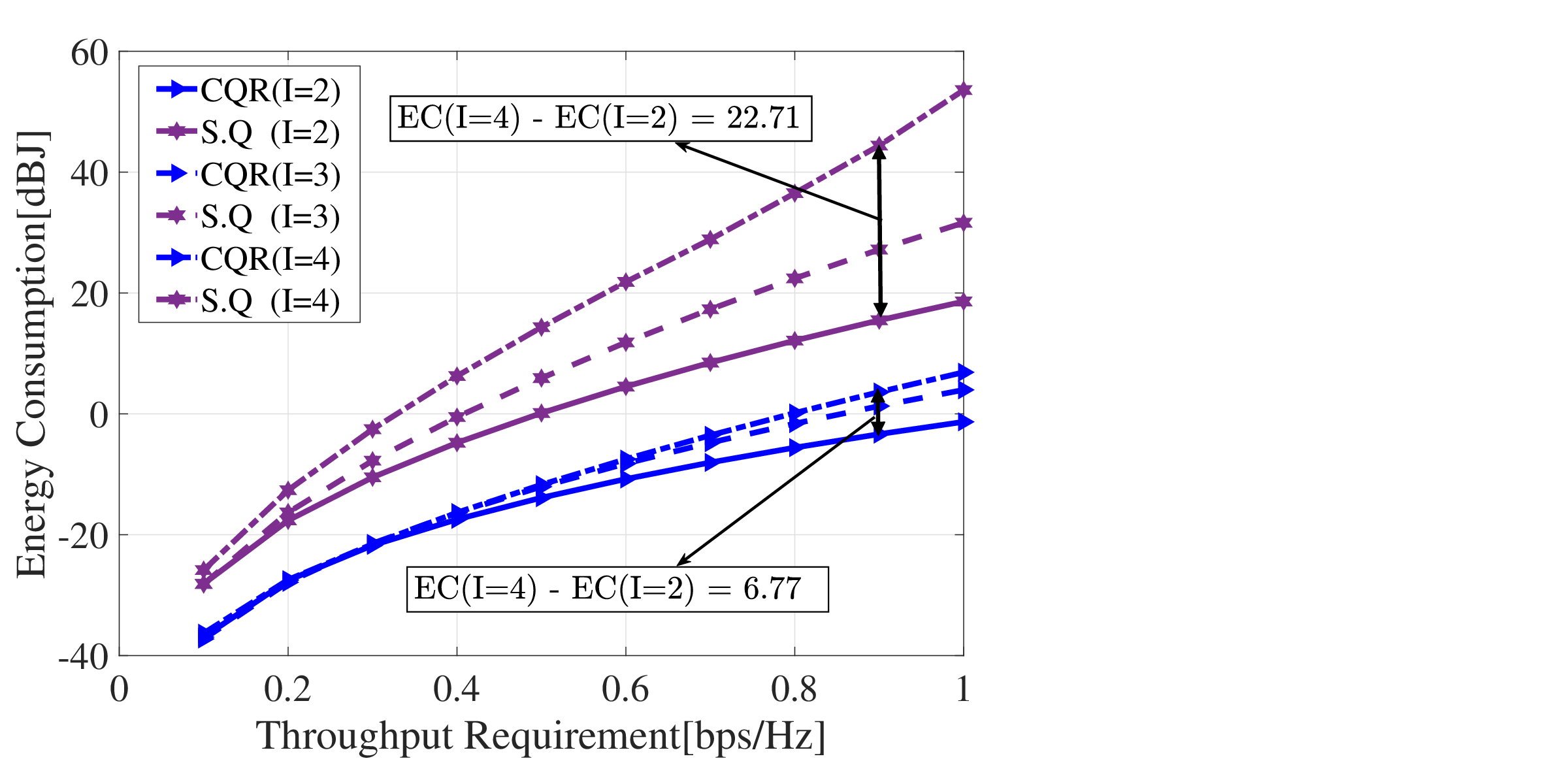}
\caption{EC versus SBDs throughput requirement\\
   where $N=1$ and $I = 2,3,4$}
\end{center}
\end{figure}

\begin{figure}
\begin{center}
\includegraphics[height=5.5cm, width=13cm]{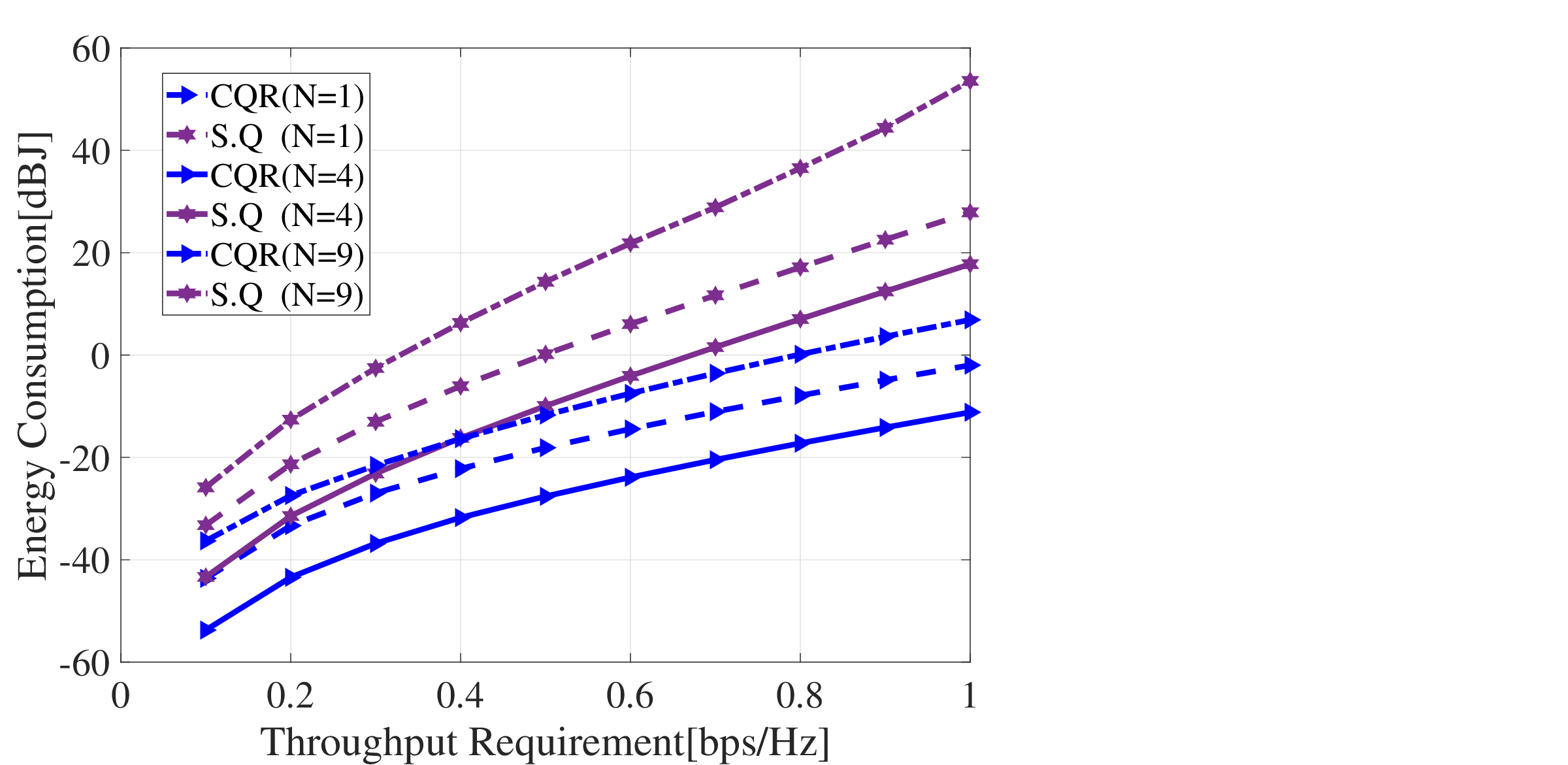}
\caption{EC versus SBDs throughput requirement \\
  where $I=4$ and ${N} = 1,4,9$}
\end{center}
\end{figure}

By narrowing the beam of the BS antenna, more energy can be directed towards the SBD, resulting in a higher SNR. As a result, ${\text{SB}}{{\text{D}}_{\text{i}}}$ can achieve the desired data rate ${C_i}$ while consuming less energy (Eq. 10a). Moreover, the increased power of the received signal at the SBD enables faster battery charging, which, in turn, reduces the time slot assigned to SBD in EHS mode (Eq. 10d) and limits the maximum energy harvested. However, this reduction also implies that the energy required to maintain the desired QoS can be obtained more quickly.

\subsection{Convergence and Computational Complexity}
The computational complexity of the SQ and CQR methods, presented in Section IV, and The corresponding plot is shown in Fig. 6. The variable parameter in this figure is the number of IoT devices (SBDs). As shown in Fig. 6, the computational complexity of the CQR method is much lower than that of the SQ method. Moreover, in the CQR method, the growth rate of computational complexity is lower than that of the SQ method with an increase in the number of SBDs in the network. Therefore, this method is suitable for implementing dense networks with a large number of users. Additionally, in the CQR method, the algorithm's complexity slightly increases as the value of $M$ increases.

\begin{figure}
\begin{center}
\includegraphics[height=5.5cm, width=13cm]{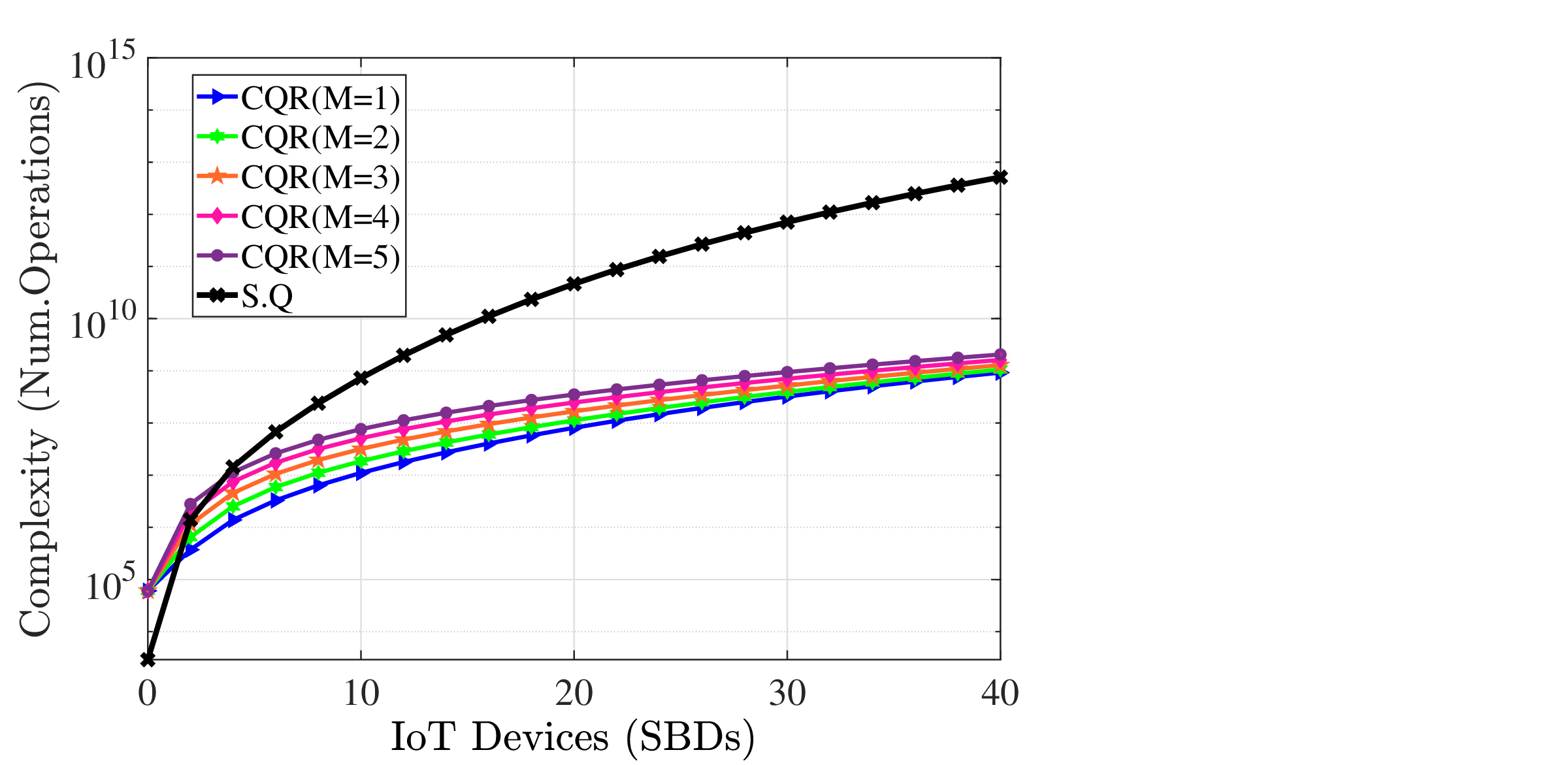}
\caption{Complexity of CQR and SQ algorithm with\\ $I=1,2,...,40$, $N = 4$ and \({\rm{\varepsilon }} = {10^{ - 6}}\)} 
\end{center}
\end{figure}

\begin{figure}
\begin{center}
\includegraphics[height=5.5cm, width=13cm]{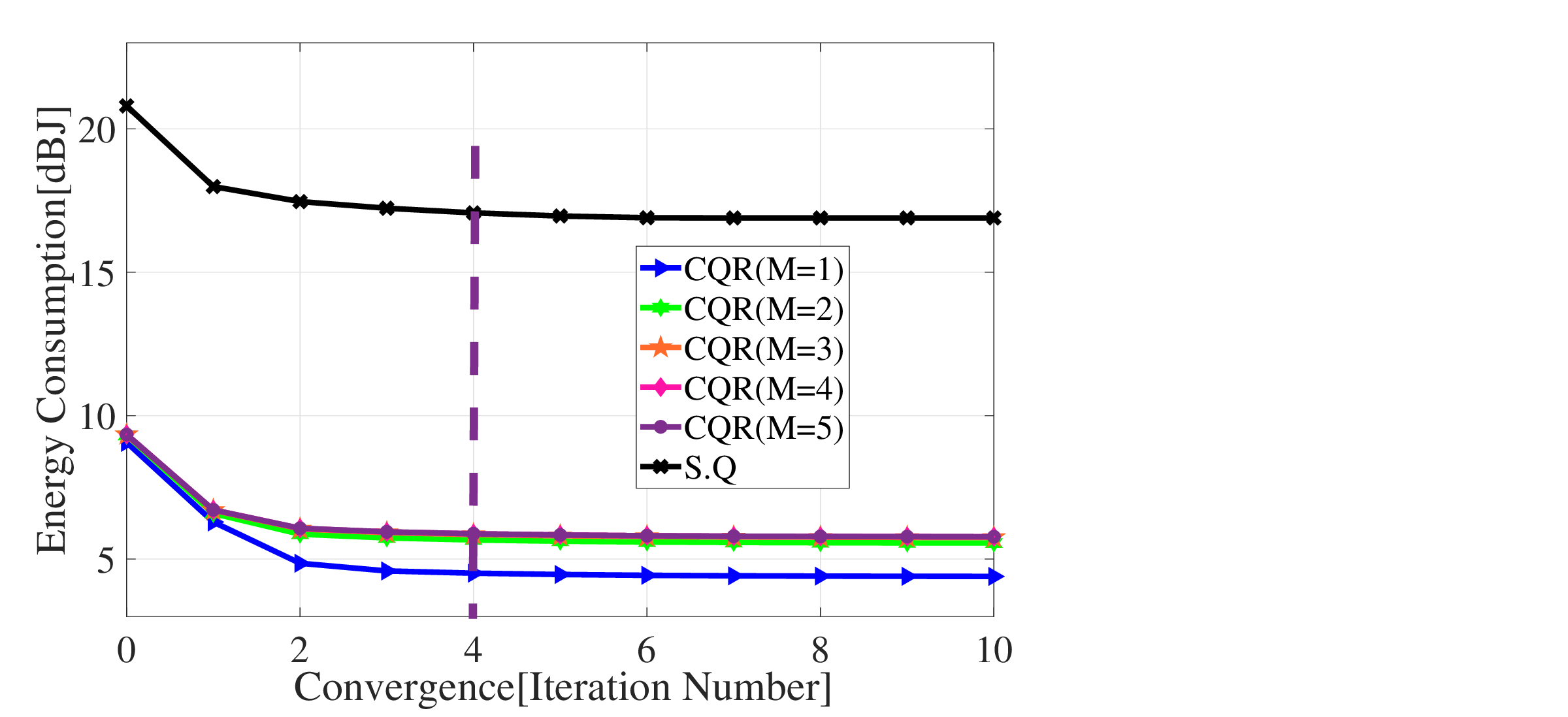}
\caption{Convergence behaviour of the CQR and SQ algorithm \\ with \({C_i=8 \,\,bps/Hz}\) and \({I=5}\)} 
\end{center}
\end{figure}
According to the description in the sections \emph{V.A}, the CQR method achieves the least computational complexity and closest convergence to other values of the function when the approximation coefficient is set to $M=4$.

Also, the convergence plots of the proposed methods are shown in Fig. 7. As it is evident, both CQR and SQ methods had good Convergence speed and converged to the desired solution after only 4 iterations. It should be noted that the execution time of each iteration is approximately 1 second.
\subsection{Comparison of T-SR and TDMA Modes}
We are conducting a study to evaluate the T-SR system, proposed and simulated to reduce EC in B5G and 6G networks. In particular, we will compare its scheduling method with that of the TDMA system, which assigns specific time slots to users for sequential access to the network.

As illustrated in Fig. 2, the T-SR mode was introduced in section II. In order to facilitate comparison, we have included a diagram of the TDMA mode in Fig. 8. The TDMA scheme, which can be seen as a simplified version of T-SR, can be modeled by applying certain simplifications to the T-SR model, as per the given definitions.
\begin{figure}
\begin{center}
\includegraphics[width=8.7cm]{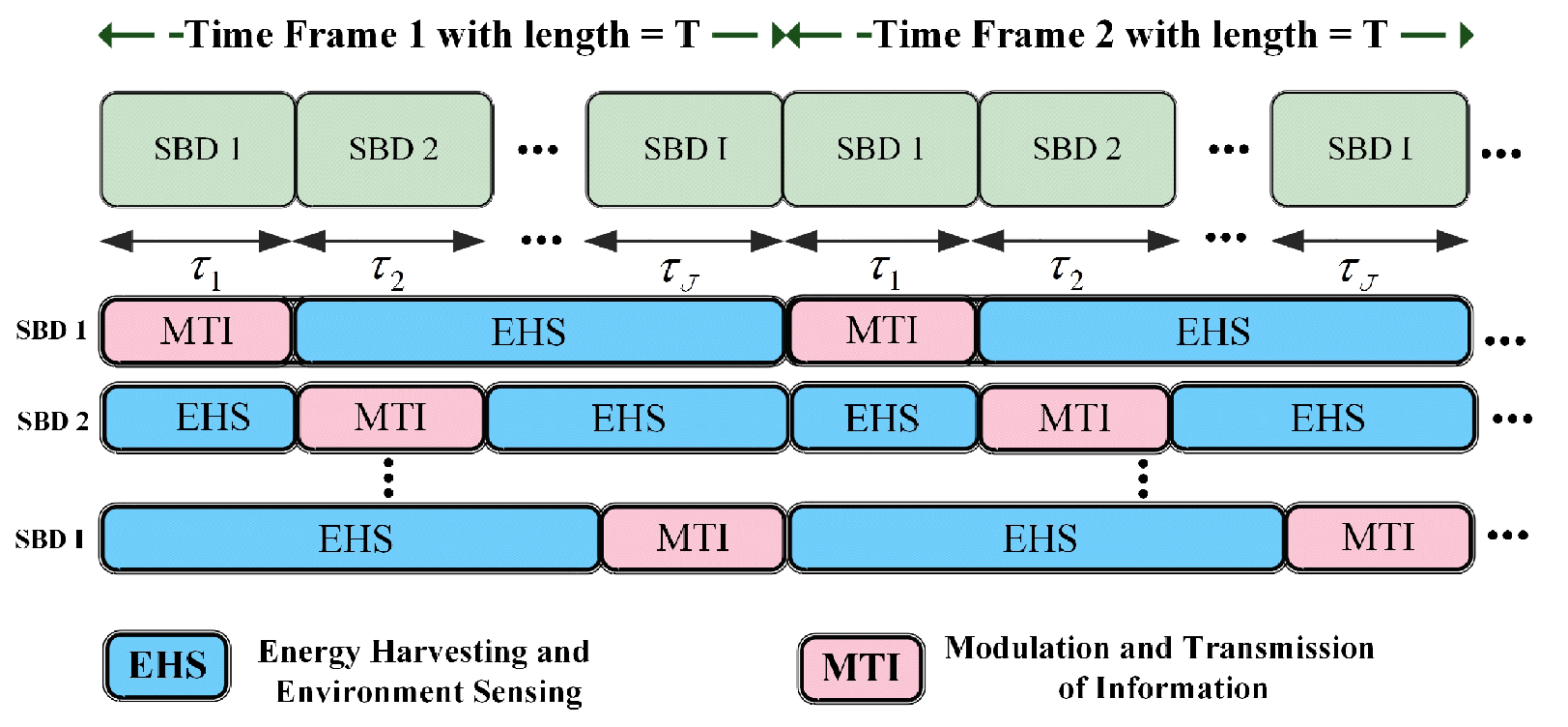}
\caption{The TDD frame for EHS and MTI modes in TDMA transmission information model} 
\end{center}
\end{figure}

\begin{figure}
\begin{center}
\includegraphics[height=5.5cm, width=13cm]{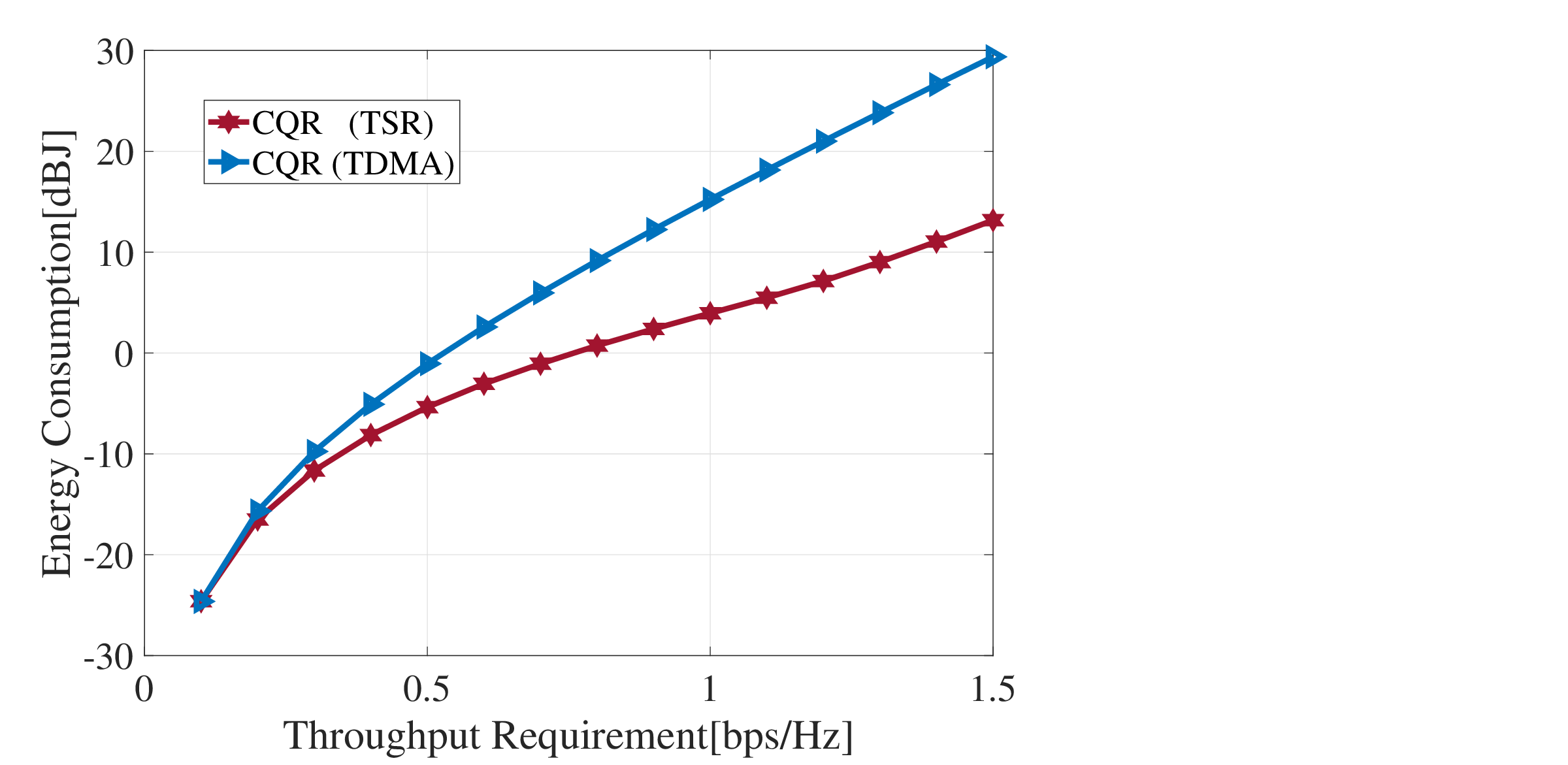}
\caption{Comparison of the EC versus SBDs throughput requirement\\ for the T-SR and TDMA scenarios, $I = 8,N = 9$} 
\end{center}
\end{figure}
In TDMA mode, each SBD takes turns transmitting its information in equal time slots within designated time frames \cite{r39}. If the SBD needs to transmit a large amount of data, it may require more time slots and multiple time frames to complete its transmission. This is because the amount of data that can be transmitted within a single time slot is limited, and larger data transmissions require more time slots, which may extend beyond a single time frame.

The TDMA mode was simulates by assuming its successful implementation in the network without any interference in the receivers. Fig. 9 presents a comparison of the EC between the T-SR mode in the SR network and the TDMA mode in typical networks. As shown in Fig. 8, the T-SR mode consumes significantly less energy than the TDMA mode (by about 8 dB), owing to the optimal allocation of time slots based on the specific needs of each SBD and the ability to transmit information and harvest energy at any time.

In TDMA mode, SBDs may transmit their data non-continuously over multiple time frames, while the T-SR mode enables the transmission of data to be completed within a single time frame using sequential time slots. The reason for this difference is that in TDMA mode, SBDs are only required to transmit their data during their designated time slot within the assigned time frame. As a result, if the SBD needs to transmit a large amount of data that cannot be accommodated within a single time slot, it must transmit the data over multiple time frames, leading to non-continuous transmission. In contrast, T-SR mode utilizes sequential time slots within a single time frame, allowing for continuous transmission of data within a single time frame.Therefore, this difference leads to more energy loss in TDMA mode compared to T-SR mode, resulting in decreased EE. This difference increases further with the increase in the number of users.

\subsection{Comparison Proposed Method With Other IoT Protocols}
As technology advances, we can expect the widespread use of IoT networks in the future. A promising method for implementing these networks is by utilizing a semi-passive structure, similar to the SR system. This approach is well-suited for facilitating communication between IoT users and upcoming cellular networks such as 6G and B5G. The use of SR systems offers several advantages, including not requiring additional infrastructure, increasing SE, and improving EE.
\begin{figure}
\begin{center}
\includegraphics[height=5.5cm, width=13cm]{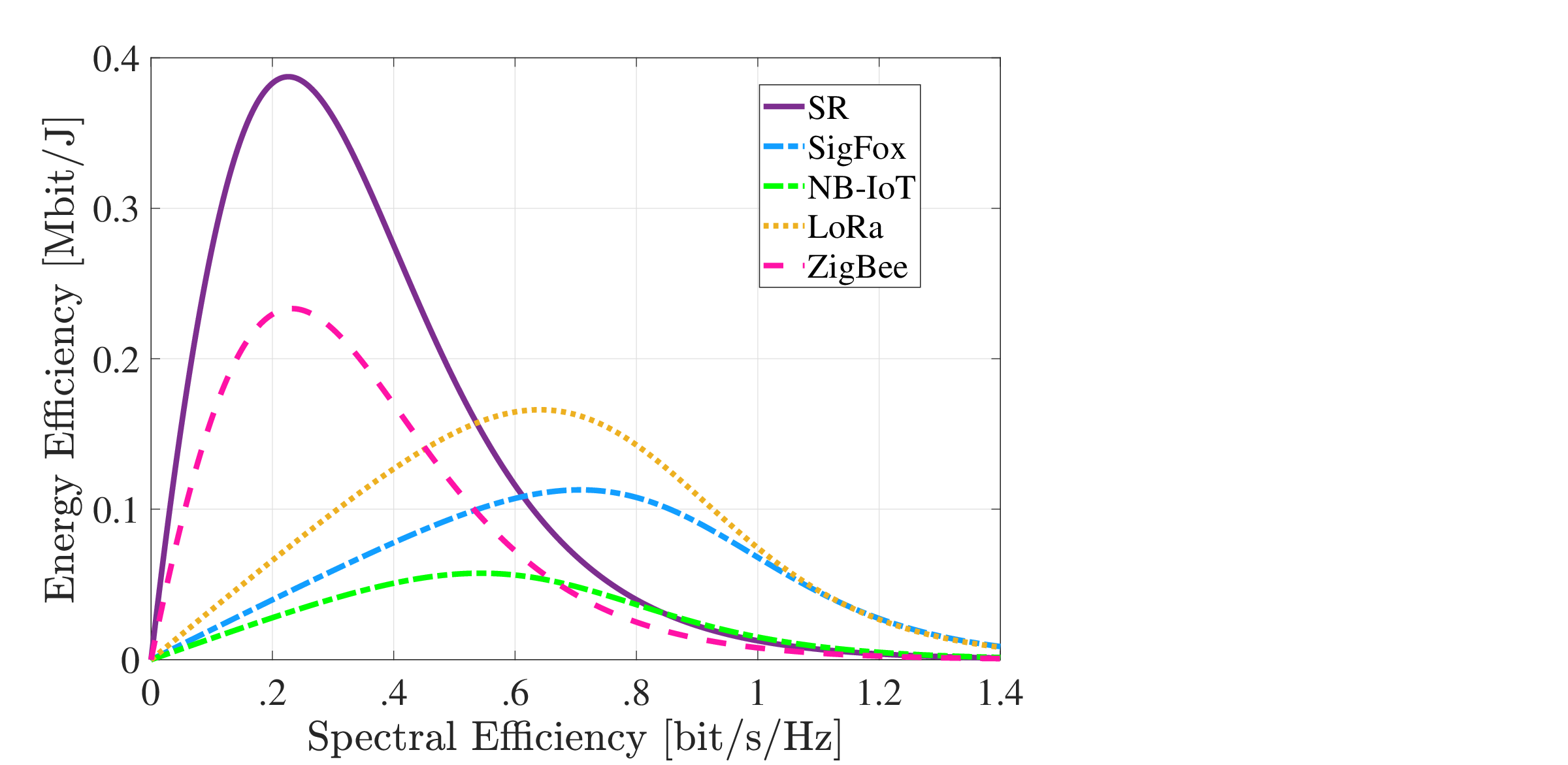}
\caption{ Energy Efficiency versus Spectral Efficiency between IoT protocols and SR system }
\end{center}
\end{figure}
\begin{table}
\centering
\caption{Power Consumption (P.C) of SR and IoT protocols for one IoT device and in one transmission slot.}
\begin{tabular}{|c|c|c|c|c|}
        \Xhline{1.2pt} 
\textbf{Protocols} & \textbf{Frequency} & \textbf{BW} & \textbf{P.C} & \textbf{Ref} \\
        \Xhline{1.2pt} 
SigFox & 902 MHz & 200 KHz & $\sim$100 mW &  \\
\cline{1-4}
LoRa & 928 MHz & 500 KHz & $\sim$150 mW & \cite{r2,r61,r62,r63,r64,r65,r66} \\
\cline{1-4}
NB-IoT & 1.8 GHz & 1 MHz & $\sim$500 mW &  \\
\hline
ZigBee & 2.4 GHz & 2 MHz & $\sim$100 mW & \cite{r67,r68} \\
\hline
SR & 2 GHz & 400 KHz & $\sim$11.6 mW & This Paper \\
\hline
\end{tabular}
\end{table}

IoT networks can be implemented using various protocols, such as sensor networks that use batteries like LoRa and ZigBee, or sensors with wireless energy harvesting in wireless powered communication networks (WPCN). These systems require active RF energy-consuming components like mixers and power amplifiers, and a complex electrical and communication infrastructure to transmit information. In contrast, the SR system utilizes a passive structure with SBDs that have extremely low EC, and leverage the existing infrastructure of other communication networks, such as cellular and Wi-Fi. As a result, the EC in the network is greatly reduced, making the SR system a more energy-efficient and cost-effective approach for implementing IoT networks. 

This section aims to compare the EE of IoT devices implemented with different systems, such as SR, ZigBee, LoRa, SigFox, and NB-IoT, irrespective of the infrastructure's EC. In this paper, we define EE as the ratio between the instantaneous throughput and the total power consumption. By comparing these systems, we can determine which one is more energy-efficient and suitable for IoT networks.

Table III provides the frequency band, bandwidth, and power consumption values for each protocol, as specified in their respective standards. The SR system is expected to be implemented within the approximate frequency band and bandwidth range outlined in the table (which corresponds to ambient waves). Fig. 10 depicts the relationship between EE and SE for each system, assuming identical conditions for all systems. We assume that all energy is used for sending and receiving information, and no energy consumption occurs in the circuit of the device. Additionally, we disregard the energy consumption associated with the infrastructure of each system. The diagram represents the transmission and reception of a single signal within a single time slot. As shown in Fig. 10, the SR system exhibits significantly better EE compared to other IoT protocols.

\subsection{The Impact of Different Locations of SBDs on Network Performance}

\begin{figure}
    \centering
\begin{center}
\includegraphics[height=5.5cm, width=13cm]{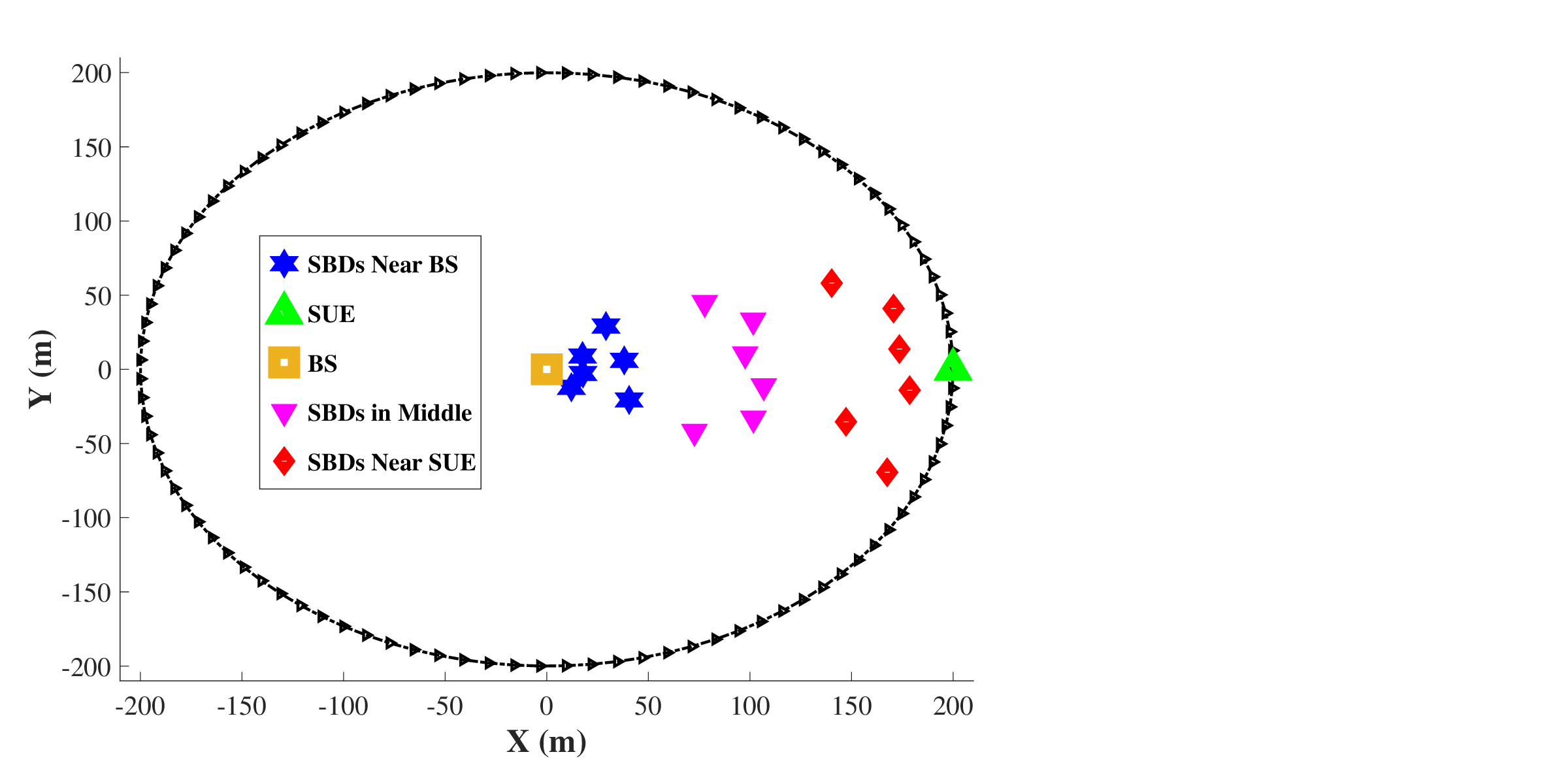}
\caption{Energy Consumption in Symbiotic radio network with change the location of SBDs} 
\end{center}
\end{figure}

\begin{figure}
    \centering
\begin{center}
\includegraphics[height=5.5cm, width=13cm]{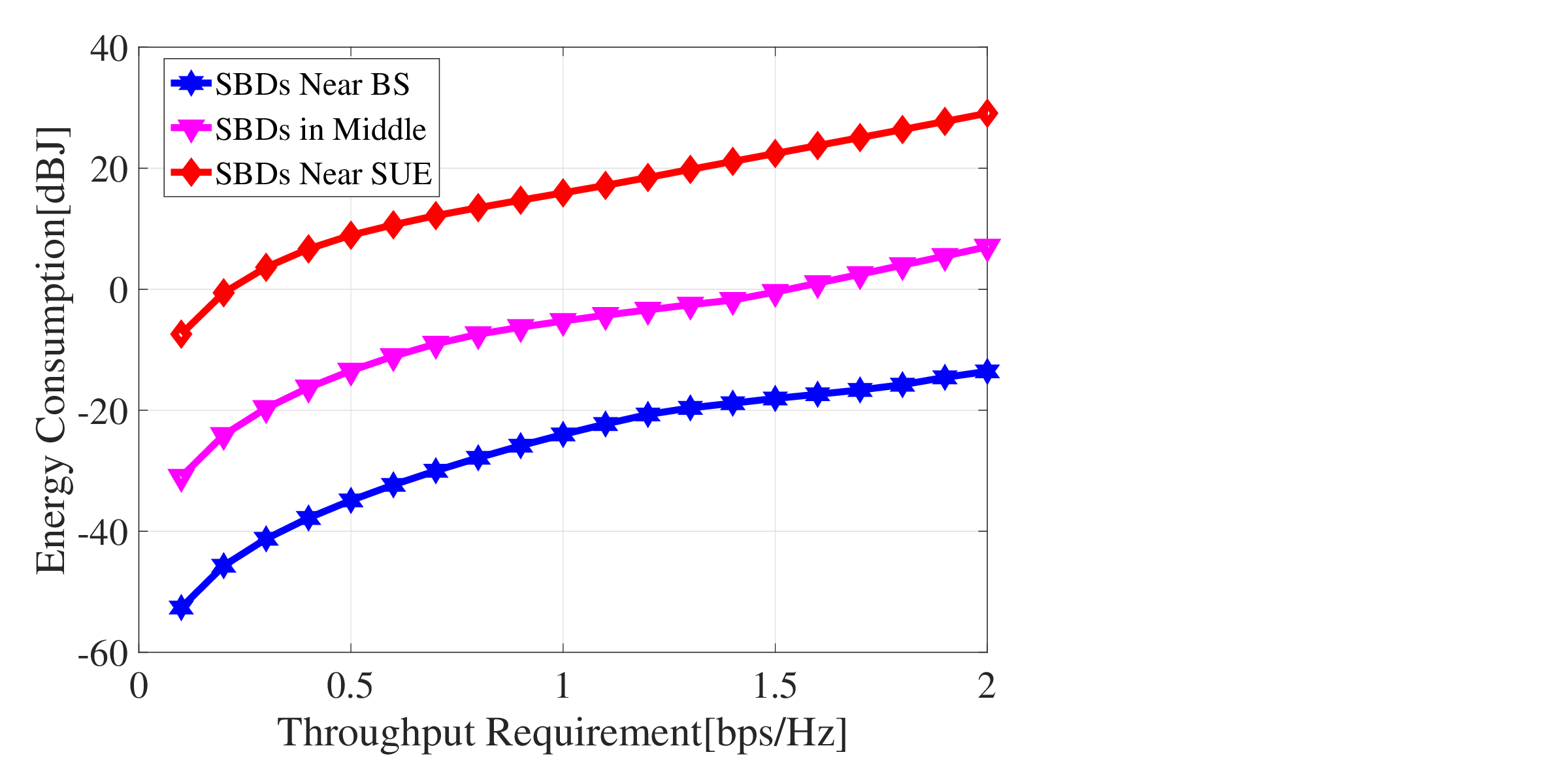}
\caption{Symbiotic radio network  topology with different Locations of SBDs} 
\end{center}
\end{figure}

We will conduct simulations using the CQR technique with $M=4$ and $I=6$, and the results can be generalized to other cases. Assuming a coverage radius of 200 meters for the BS and the SUE being located at the maximum distance from the BS. We consider the SR network topology as shown in Fig. 11, where three scenarios arise: 1) SBDs located near the BS, 2) SBDs located in the middle space between the BS and SUE, and 3) SBDs located near the SUE. As shown in Fig. 12, we observe that in the assumed SR network, the closer the SBDs are to the BS, the lower the energy consumption in the network, while the closer they are to the SUE, the higher the energy consumption. This is because as SBDs get closer to the BS, they receive a stronger signal and the channel gain, which is related to the square inverse of the distance between BS-SBDs, also increases, resulting in an increased SNR received by SBD. Thus, based on constraints (10a), ${\text{SB}}{{\text{D}}_{\text{i}}}$ achieves the desired rate ${C_i}$ with less energy. Additionally, due to the reduction in the timeslot assigned in EHS mode to SBD, the maximum energy harvested (constraint 10d) by it decreases as the energy required for transmitting information is completed faster. This ultimately results in decreased energy consumption in the entire network, as seen when SBDs are close to the BS.

Conversely, when SBDs are close to the SUE, the received signal power and their SNR decrease, leading to a higher energy consumption required to achieve the desired QoS in the system.

\section{CONCLUSION}
 In this paper, we consider a novel solution in the SR network with CSR setup for achieving energy efficiency network. When SBDs have data to transmit, they modulate the information onto the received ambient RF signal and send it to the intended SUE. The main objective of this paper is to enhance energy efficiency in this network by minimizing energy consumption while ensuring the minimum required throughput for SBDs. To achieve this, we propose a new scheduling scheme called T-SR that optimally allocates resources to SBDs. We formulate the T-SR system as a non-convex optimization problem. To solve this problem, we use mathematical techniques and introduce a new approach called SQ and CQR to relax and discipline the problem, which reduces its complexity and convergence time.
 
In the simulation section, we compare the performance of the CQR and SQ methods and observe that the CQR method yields better EE. Furthermore, by changing basic network parameters such as the number of IoT devices and the number of active massive MIMO antennas in the BS, CQR was stable and could further reduce EC in the network. Additionally, we compare the proposed system with the traditional TDMA system and find that the SR system with TDMA mode achieves better EE. Furthermore, Fig. 10 shows that the power consumption of the network is much lower when using the SR system for communication among a large number of IoT users compared to other current IoT protocols. These findings suggest that the proposed system has significant potential for reducing EC in future-generation networks on a global scale.

{\appendix}

We first obtain the matrix ${\bf{A}}$ which satisfies the following relation for \(\left( {{\theta _i},{\tau _j}} \right)\):
{\small\begin{equation}
\left[ {\begin{array}{*{20}{c}}
{\frac{{{\tau _j}}}{{{{\left( {{\tau _j} + {\theta _i}} \right)}^2}}}}&{ - \frac{{{\theta _i}}}{{{{\left( {{\tau _j} + {\theta _i}} \right)}^2}}}}\\
{ - \frac{{{\theta _i}}}{{{{\left( {{\tau _j} + {\theta _i}} \right)}^2}}}}&{\frac{{\theta _i^2}}{{{\tau _j}{{\left( {{\tau _j} + {\theta _i}} \right)}^2}}}}
\end{array}} \right] \le {\bf{A}} = \left[ {\begin{array}{*{20}{c}}
{{{\rm{a}}_{{\rm{11}}}}}&{{{\rm{a}}_{{\rm{12}}}}}\\
{{{\rm{a}}_{21}}}&{{{\rm{a}}_{{\rm{22}}}}}
\end{array}} \right]
\end{equation}}
According to Eq.(33) the largest value of \({{\rm{a}}_{{\rm{11}}}}\) is obtained when \({\theta _i}\) and ${\tau _j}$ have their lowest values
{\small\begin{equation}
{a_{11}} = \mathop {\arg \max }\limits_{{\theta _i},{\tau _j}} \left\{ {{{{\tau _j}} \mathord{\left/
 {\vphantom {{{\tau _j}} {{{\left( {{\tau _j} + {\theta _i}} \right)}^2}}}} \right.
 \kern-\nulldelimiterspace} {{{\left( {{\tau _j} + {\theta _i}} \right)}^2}}}} \right\}\left| \begin{array}{l}
{\theta _i} = 0\\
{\tau _j} = {{{{\hat \tau }_j}} \mathord{\left/
 {\vphantom {{{{\hat \tau }_j}} \beta }} \right.
 \kern-\nulldelimiterspace} \beta }
\end{array} \right. = \frac{\beta }{{{{\hat \tau }_j}}}
\end{equation}}

To obtain the maximum value of \({{\rm{a}}_{{\rm{22}}}}\) based on the function \(\frac{{\theta _i^2}}{{{\tau _j}{{\left( {{\tau _j} + {\theta _i}} \right)}^2}}}\) is monotonically increasing with respect to \(\theta _i^{}\) and monotonically decreasing with respect to ${\tau _j}$:
{\small\begin{equation}
{a_{22}} = \mathop {\arg \max }\limits_{{\theta _i},{\tau _j}} \left\{ {\frac{{\theta _i^2}}{{{\tau _j}{{\left( {{\tau _j} + {\theta _i}} \right)}^2}}}} \right\}\left| \begin{array}{l}
\mathrm{Lim}\,{\theta _i} \to \infty \\
{\tau _j} = {{{{\hat \tau }_j}} \mathord{\left/
 {\vphantom {{{{\hat \tau }_j}} \beta }} \right.
 \kern-\nulldelimiterspace} \beta }
\end{array} \right. = \frac{\beta }{{{{\hat \tau }_j}}}
\end{equation}}

To obtain \({{\rm{a}}_{21}} = {{\rm{a}}_{{\rm{12}}}}\), the derivative of the function \(\frac{{{\theta _i}}}{{{{\left( {{\tau _j} + {\theta _i}} \right)}^2}}}\) is calculated and it is observed that the maximum value of this function is obtained for $\tau _i^{} = \theta _i^{}$ and also $\tau _i^{}$ has it's lowest value, so: 
{\small\begin{equation}
{a_{12}} = {a_{21}} \le \mathop {\arg \max }\limits_{{\theta _i},{\tau _j}} \left\{ {\frac{{{\theta _i}}}{{{{\left( {{\tau _j} + {\theta _i}} \right)}^2}}}} \right\}\left| \begin{array}{l}
\,{\theta _i} = {{{{\hat \tau }_j}} \mathord{\left/
 {\vphantom {{{{\hat \tau }_j}} \beta }} \right.
 \kern-\nulldelimiterspace} \beta }\\
{\tau _j} = {{{{\hat \tau }_j}} \mathord{\left/
 {\vphantom {{{{\hat \tau }_j}} \beta }} \right.
 \kern-\nulldelimiterspace} \beta }
\end{array} \right. =\frac{\beta }{{{{4\hat \tau }_i}}}
\end{equation}}

In addition, to satisfy the above relations, the determinant of the matrix (47) should be greater than zero. By defining an auxiliary variable $\Re $, this inequality will be simplified as follows:
 {\small\begin{equation}
\begin{array}{*{20}{l}}
{\left( \begin{array}{l}
\underbrace {\left( {{{\rm{a}}_{{\rm{11}}}} - \frac{{{\tau _j}}}{{{{\left( {{\tau _j} + {\theta _i}} \right)}^2}}}} \right)}_{{\Re _1}}\underbrace {\left( {{{\rm{a}}_{{\rm{22}}}} - \frac{{\theta _i^2}}{{{\tau _j}{{\left( {{\tau _j} + {\theta _i}} \right)}^2}}}} \right)}_{{\Re _2}}\\
 - \underbrace {{{\left( {{{\rm{a}}_{{\rm{12}}}} + \frac{{{\theta _i}}}{{{{\left( {{\tau _j} + {\theta _i}} \right)}^2}}}} \right)}^2}}_\Re 
\end{array} \right)}
\end{array} \ge 0
\end{equation}}
Consequently, based on Eq.(51) we calculate the \({{\rm{a}}_{21}} = {{\rm{a}}_{{\rm{12}}}}\):
{\small\begin{equation}
\begin{array}{l}
{{\rm{a}}_{{\rm{12}}}} = {a_{21}} =  - \frac{1}{2}\left( {\max \left( {\frac{{{\theta _i}}}{{{{\left( {{\tau _j} + {\theta _i}} \right)}^2}}}} \right) - \min \left( {\frac{{{\theta _i}}}{{{{\left( {{\tau _j} + {\theta _i}} \right)}^2}}}} \right)} \right)\\
 \,\,\,\,\,\,\,\,\,\, =  - \frac{1}{2}\left( {\frac{\beta }{{4{{\hat \tau }_j}}} - 0} \right) =  - \frac{\beta }{{8{{\hat \tau }_j}}}
\end{array}
\end{equation}}

The following inequality is established for \(\Re \) :
{\small\begin{equation}
\Re  \ge \left( { - \frac{\beta }{{8{{\hat \tau }_j}}} + \frac{\beta }{{4{{\hat \tau }_j}}}} \right) = \frac{\beta }{{8{{\hat \tau }_j}}}
\end{equation}}

and also, we have the following relationships:
{\small\begin{equation}
{{\rm{a}}_{{\rm{11}}}} \le {\Re _1} + \frac{\beta }{{{{\hat \tau }_j}}} = \frac{{9\beta }}{{8{{\hat \tau }_j}}}\,\,{\rm{,}}\,{{\rm{a}}_{{\rm{22}}}} \le {\Re _2} + \frac{\beta }{{{{\hat \tau }_j}}} = \frac{{9\beta }}{{8{{\hat \tau }_j}}}
\end{equation}}

Finally, after these calculations, an upper bound of matrix (32) is shown by Eq.(34).
\\
{\small\bibliographystyle{IEEEtran}
\bibliography{phdrefs}}

\begin{IEEEbiography}[{\includegraphics[width=1in,height=1.25in,clip,keepaspectratio]{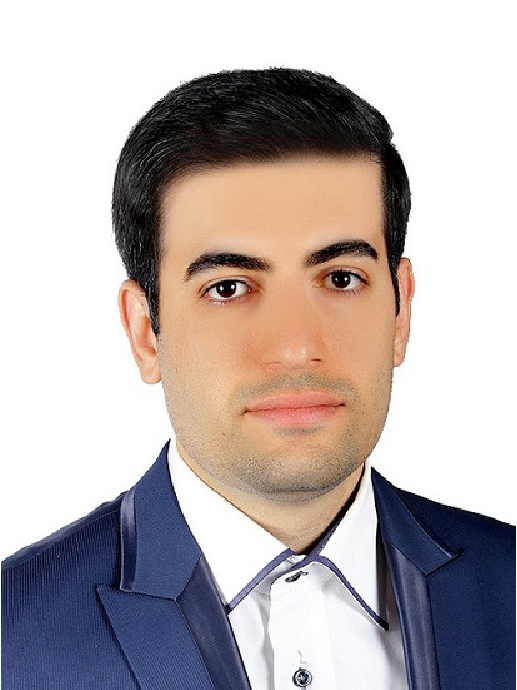}}]{Rahman Saadat Yeganeh}
received the M.Sc. degree in electrical engineering (telecommunication systems) from the University of SRBIAU, Tehran, Iran in 2017. He is currently a Ph.D. candidate in electrical engineering (telecommunication systems) at the Isfahan University of Technology (IUT), Isfahan, Iran. He is also a researcher involved in the design of telecommunication systems. His research interests include wireless communication systems, wireless power transfer, symbiotic radio, reconfigurable intelligent surfaces, beamforming, NOMA, and signal processing.
\end{IEEEbiography}
\vspace{-1.2cm}
\begin{IEEEbiography}[{\includegraphics[width=1in,height=1.25in,clip,keepaspectratio]{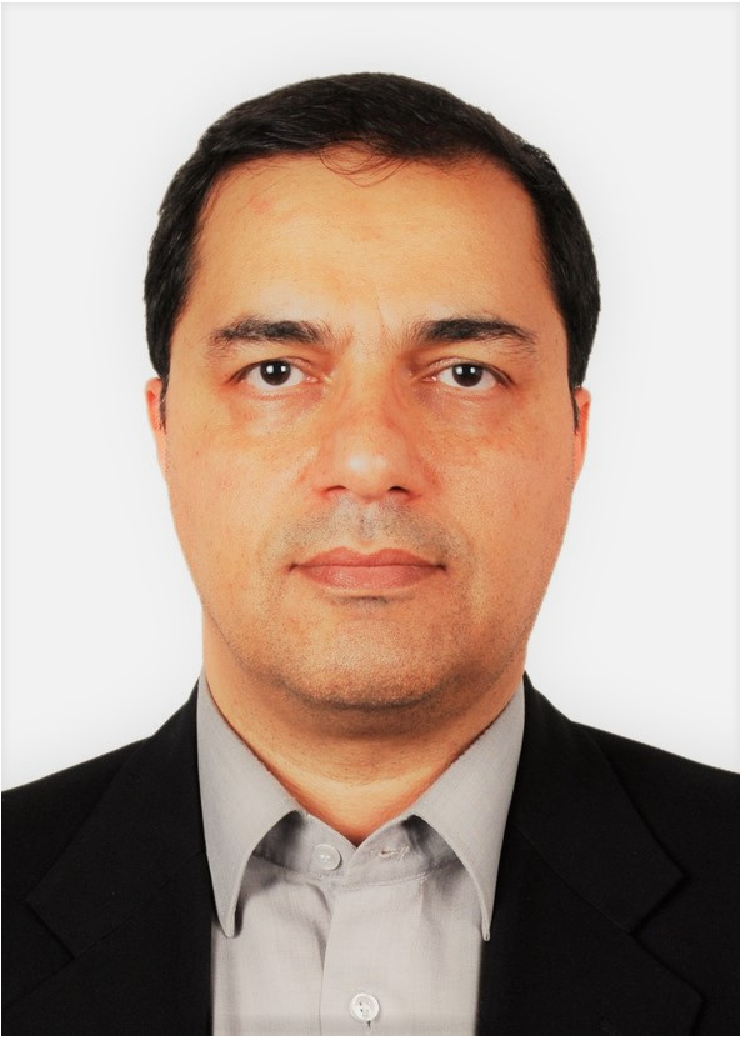}}]{Mohammad Javad Omidi}
received his Ph.D. from University of Toronto in 1998. He worked in industry by joining a research and development group designing broadband communication systems for 5 years in US and Canada. In 2003 he joined the Department of Electrical and Computer Engineering, at Isfahan University of Technology (IUT), Iran; and then served as the chair of Information Technology Center and chair of the ECE department at this university. Currently he is the Director of IRIS a UNESCO organization in Iran, and the VP for Research and Development at Isfahan Science and Technology Town. He has numerous publications and more than 15 US and international patents in the area of telecommunications. His scientific research interests are in the areas of mobile computing, wireless communications, digital communication systems, software radio, cognitive radio, and VLSI architectures for communication algorithms.
\end{IEEEbiography}
\vspace{-1.2cm}
\begin{IEEEbiography}[{\includegraphics[width=1in,height=1.25in,clip,keepaspectratio]{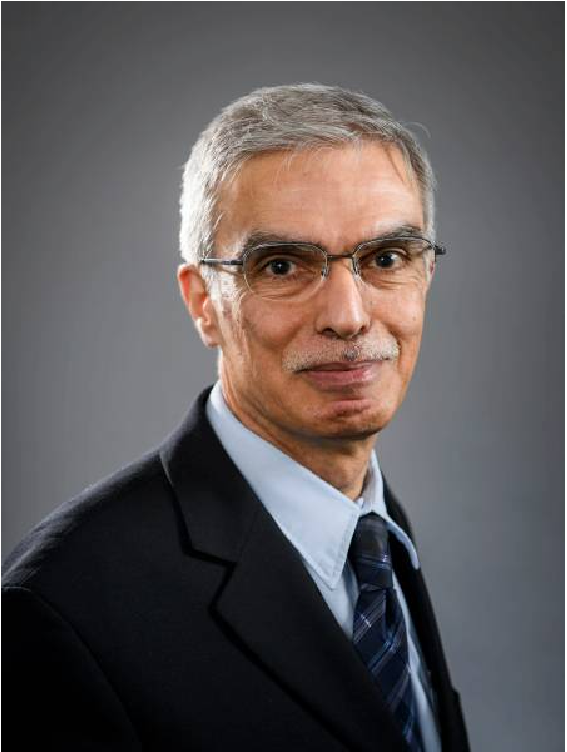}}]{Mohammad Ghavami}
(Senior Member, IEEE) is currently a Professor in telecommunications at London South Bank University. Prior to this appointment, he was with King’s College London from 2002 to 2010 and the Sony Computer Science Laboratories in Tokyo from 2000 to 2002. He has authored the books \emph{Ultra Wideband Signals and Systems in Communication Engineering} and \emph{Adaptive Antenna Systems}. He has also published over 180 technical papers mainly related to UWB and its medical applications. He holds three U.S. and one European patents. He won the esteemed European Information Society Technologies Prize in 2005 and two invention awards from Sony. He has been the Guest Editor of the \emph{IET Proceedings Communications} Special Issue on Ultra Wideband Systems and the Associate Editor of the Special Issue of the \emph{IEICE Journal on UWB Communications}.
\end{IEEEbiography}

\end{document}